\documentclass[a4paper]{article}
\usepackage[pdftex]{graphicx}
\pdfoutput=1
\usepackage{amsmath}
\usepackage{amssymb}
\usepackage{amsthm}
\usepackage{fullpage}
\newcommand{\abs}[1]{\left|#1\right|}
\usepackage[utf8]{inputenc}
\usepackage[T1]{fontenc}
\usepackage{hyperref}
\usepackage{todonotes}

\newtheorem{definition}{Definition}

\newcount\colveccount
\newcommand*\colvec[1]{
        \global\colveccount#1
        \begin{pmatrix}
        \colvecnext
}
\def\colvecnext#1{
        #1
        \global\advance\colveccount-1
        \ifnum\colveccount>0
                \\
                \expandafter\colvecnext
        \else
                \end{pmatrix}
        \fi
}
\usepackage{epstopdf}
\usepackage{subcaption}

\title{Imperfections, impacts, and the singularity of Euler's disk}
\author{Tam\'as Baranyai and P\'eter L. V\'arkonyi}

\begin{document}

\maketitle

\begin{abstract}
The motion of a rigid, spinning disk on a flat surface ends with a dissipation-induced finite-time singularity. The problem of finding the dominant energy absorption mechanism during the last phase of the motion generated a lively debate during the past two decades. Various candidates including air drag and various types of friction have been considered, nevertheless impacts have not been examined until now. We investigate the effect of impacts caused by geometric imperfections of the disk and of the underlying flat surface, through analysing the dynamics of polygonal disks with unilateral point contacts. Similarly to earlier works, we determine the rate of energy absorption under the assumption of a regular pattern of motion analogous to precession-free motion of a rolling disk. In addition, we demonstrate that the asymptotic stability of this motion depends on parameters of the impact model. In the case of instability, the emerging irregular motion is investigated numerically. We conclude that there exists a range of model parameters (small radii of gyration or small restitution coefficients) in which absorption by impacts dominates all preiously investigated mechanisms during the last phase of motion. Nevertheless the parameter values associated with a homogenous disk on a hard surface are typically not in this range, hence the effect of impacts is in that case not dominant.
\end{abstract}

\section{Introduction}

Euler's disk is a popular scientific toy. Its dynamic behaviour captured the attention of many scientists over the past decades. 
%
Euler's disk consists of a flat disk, which is spun over a hard, slightly convace surface similarly to a spinning coin on a table. Its spinning motion often goes on for several minutes, while its mechanical energy is slowly dissipated. During the final phase of the motion, the inclination angle $\alpha$ of the disk approaches 0, while the rate of spinning appears to diverge towards inifnity. The motion is accompanied by loud noise whose frequency also appears to approach infinity until the disk abruptly stops with one of its faces resting on the table.  

Attempts to identify the dominant energy absorption mechanism responsible for the finite-time singularity of the spinning disk generated a lively debate among researchers. This was initiated by a letter of H. K. Moffatt  to Nature in 2000 \cite{moffatt2000euler}, in which he argued that the dominant absorption mechanisms during the last phase of motion is viscous damping of the thin air layer between the coin and the underlying surface. In a followup letter,Van den Engh et al. \cite{van2000analytical} argued that \textit{slipping friction} caused by small-amplitude precessional motion of the disk (not visible to the bare eye) is the dominant factor instead.  This was followed by a series of other papers and notes  involving critical treatment of modeling assumptions \cite{ruina2000unp}\cite{bildsten2002viscous}; analytical investigations   \cite{le2005dynamics} \cite{leine2009experimental}; numerical simulations \cite{le2005dynamics} \cite{kessler2002ringing} \cite{saje2006rattling}; and experiments \cite{easwar2002speeding}  \cite{mcdonald2000rolling} \cite{leine2009experimental} \cite{caps2004rolling} \cite{petrie2002does} (including ones in vacuum \cite{van2000analytical}). A review of this body of works was published by Leine \cite{leine2009experimental}, who analyzed the effect of various models of \textit{rolling friction} as well as of air drag in a unified approach. He concluded that dry and viscous \textit{rolling friction} are likely to govern the motion during the last few seconds of the motion, and the effect of air drag is likely to become dominant over friction only during the last milliseconds. His analysis was based on the assumption of precession-free motion and did not investigate how precession effects energy dissipation. 


The ideally dissipation-free dynamics of a rigid, rolling disk has a one-parameter family of precession-free, "steady" solutions, for which the inclination of the disk remains constant and its center of mass remains immobile \cite{o1996dynamics}  \cite{kuleshov2001steady} \cite{paris2002disk} \cite{batista2006steady}. On the short run, the observed motion of Euler's disk appears to mimic one of these solutions. In line with this observation, the theoretical analysis of the singularity is often based on the assumption that the disk slowly drifts along this one-parameter family of solutions as dictated by the rate of energy dissipation \cite{leine2009experimental}. This model seems plausible due to the low rate of dissipation in the system, except at the very end of the motion when the dissipation rates of certain mechanisms blow up according to theoretical models. 
This kind of analysis typically predicts energy profiles of the form
\begin{align}
E(t)=a(t_f-t)^c
\label{eq:scalinglaw}
\end{align}
where $t$ is time, $E$ is the total mechanical energy of the system, $t_f$ is the time of the singularity, and $a,c$ are positive constants where the exponent $c$ depends on the type of dissipation mechanism. Clearly, the dissipation mechanism with the lowest exponent $c$ will dominate over all others at the end of the mtion. It was found that the exponent associated with air drag is $1/2$ \cite{moffatt2000euler} or $4/9$ \cite{bildsten2002viscous} (the latter is based on an improved model of boundary layer effects) and the exponents associated with various models of rolling friction are $1/2$, $2/3$ or $2$. In addition, some models of friction create asymptotic convergence of $E$ to 0 rather than a finite-time singularity (which largely corresponds to $c=\infty$). 

Other works focused on qualitative differences between the actual motion of a disk and the idealized picture sketched above. The steady motion of the dissipation free disk corresponds to a (Lyapunov stable) center of the dynamics in a rotating frame. The works \cite{stanislavsky2001nonlinear} \cite{saje2006rattling} point out that the frequency of small-amplitude oscillations around this center (i.e. precessional motion) matches the frequency of the audible noise produced by the disk. Numerical simulation  and experiments both confrm the presence of the precessional component.   \cite{caps2004rolling}, \cite{kessler2002ringing}. In addition, \cite{villanueva2005vibrations} also showed that the disk may slide during its motion (but see a somewhat controversary result in  \cite{petrie2002does}). The recent paper \cite{borisov2015loss} shows via phyisical experiments that it looses and reestablishes contact with the ground, whereas several earlier works \cite{ruina2000unp} 
\cite{kessler2002ringing} \cite{villanueva2005vibrations}   
  formulated the hypothesis that the onset of lift-off and impacts are responsible for the abrupt halt at the end of the motion. Nevertheless, the effect of precession, sliding, and impacts on the energy absorption rates has not been examined systematically so far.

This paper examines how contact detachment and impacts caused by geometric imperfections contribute to energy absorption near the singularity of the disk and how they affect precessional motion. Impacts may arise as a consequence of elastic vibrations \cite{kessler2002ringing} \cite{villanueva2005vibrations} but they also emerge in an ideally rigid model if either the shape of the disk or that of the ground surface is imperfect. Importantly, as we approach the singularity of the motion and the rate of spinning increases, even the slightest vibration or imperfection (or alternatively an arbitrarily small dust particle between the disk and the ground) will lead to lift-off and impacts. Impacts absorb energy, hence we expect that they have important effect on the energy balance of motion.

We will use a thin, regular $n$-gon shaped rigid disk as a conceptual model of an imperfect rolling disk. The rotational symmetry of this model will simplify the analysis, while we believe that the assumed symmetry of the imperfection does not alter the results qualitatively. For simplicity, we assume frictionless contact interaction in line with the observation that dissipation-free rolling motion does not require frictional forces in the limit of vanishing inclination angle $\alpha$ \cite{leine2009experimental}. We will define the "self-similar motion" of the system in the case of $\alpha<<1$, which is an analogue of the "steady rolling motion" of a dissipation-free, perfect rolling disk. Nevertheless self-similar motion incorporates energy dissipation via impacts, and it naturally leads to a scaling law similar to \eqref{eq:scalinglaw} (but due to the discrete nature of impacts, $E(t)$ will now be a piecewise constant function). As we will see, the exponent associated with self-similar motion is $c=2$, which predicts that energy loss due to impacts is not dominant when the disk approaches its singularity. In addition, we will also examine the stability of the self-similar motion by looking at eigenvalues of its linearized Poincaré map. Interestingly, we find that the self-similar motion may be asymptotically stable or unstable depending on choice of the impact model. In the latter case, the disk will start wobbling. Hence, the main contribution of this part of the paper is to provide a possible explanation of why the spinning motion of the disk tends to be accompanied by small-amplitude wobbling.

Next we also examine the effect of imperfections in the flat ground and the flat facets of the disk. Generically, if two nearly flat, rigid surfaces are pushed against each other, they will have 3 points of contact. Thus, we hypothesize that the last phase of the motion (with $\alpha$ very close to 0) will be \textit{bouncing (or rolling) on 3 point contacts}. To understand motion with 3 point contacts in the presence of gravity, we revisit existing results about the bouncing motion of a rod on a flat ground (with only two possible contact points). We find that the energy profile of a bouncing rod is again similar to \eqref{eq:scalinglaw} with  
$0\leq c \leq 2$. In particular, $c$ is a monotonically increasing function of the coefficient of restitution of impacts and of the radius of gyration. It may take values arbitrarily close to zero. We then demonstrate via a series of numerical simulation that bouncing on 3 contacts results in similar energy dissipation profile. We do not present analytical proof of this conclusion because 3 contact points appear to give rise to complex sequences of impacts, which can not be analyzed via the simple approach used in the case of self-similar motion. The main conclusion of this part is that for certain combinations of the model parameters, $c$ may go well below $4/9$ in which case impacts become the dominant energy absorption mechanism of the disk as it approaches its singularity.

\section{Mechanical model of the imperfect disk}
\subsection{Notation and kinematics}

Consider a "global" orthogonal coordinate frame $[ \ ]^g$ spanned by the pairwise ortogonal unit vectors $u_X$, $u_Y$ and $u_Z$. We will denote any vector expressed in this global frame by an upper index $*^g$. The horizontal ground $\mathcal{P}$ is in the plane spanned by $u_X,u_Y$. The vector $u_Z$ points vertically upwards. The acceleration of gravity is thus $-gu_Z$ where $g$ is the gravitational constant.

We also consider a "local" coordinate frame $[ \ ]^l$ fixed to the moving disk $\mathcal{B}$. It is spanned by the unit vectors $u_x$, $u_y$, $u_z$ with its origin at the center of mass $r_c$ of $\mathcal{B}$. Any vector expressed in this frame will be denoted by upper index $*^l$. For example, we always have $r_c^l=0$.

Our model of the imperfect disk is an infinitely thin regular $n$-gon (large $n$) with vertices 
\begin{align}
r^l_i=\colvec{3}{\cos(2i\pi/n)}{\sin(2i\pi/n)}{0}, i\in \{0,1...n-1\} \label{eq:koord}
\end{align} 
The position vectors of vertices (as well as any other vector) can be expressed in$[ \ ]^g$ as
\begin{align}
r_i^g=r_c^g+H_{l,g} r_i^l. \label{eq:pontoshelyzet} 
\end{align}   
where $H_{l,g}=[u_x^g,u_y^g,u_z^g]$ is a rotation matrix. The velocity $v_i$ of point $i$ with respect to $[ \ ]^g$ is expressed as
\begin{align}
v_i^g=v_c^g+\omega^g \times H_{l,g} r_i^l. \label{eq:pontossebesseg} 
\end{align}  
where $\omega$ is the angular velocity of $\mathcal{B}$ and $v_c$ is the velocity of the center of mass.


Let $m$ and $\theta$ denote the mass and the moment of inertia tensor of $\mathcal{B}$. The infinitesimal thickness and the rotational symmetry of $\mathcal{B}$ implies that $\theta$ takes the form
\begin{align}
\theta^l=m
\begin{bmatrix}
\rho^2 & 0 & 0\\
0 & \rho^2 &0\\
0 & 0 & 2\rho^2
\end{bmatrix}
\end{align}
where $\rho\approx 1/2$ if $\mathcal{B}$ is homogenous and $n>>1$.


\subsection{Continuous dynamics}

Assume that the object is subject to gravity as well as frictionless contact forces $\eta_i u_Z$ at points $r_i$. Then its acceleration and angular acceleration are determined by the Newton-Euler equations:
\begin{align}
\left(-mg+\sum_i\eta_i\right)u_Z=m\frac{d v_c}{d\tau} \label{eq:rollF} \\ 
\sum_i\left((r_i-r_c)\times \eta_i u_Z\right ) =\theta \frac{d \omega}{d\tau}+\omega \times \theta \omega \label{eq:rollM} 
\end{align} 
where $\tau$ denotes time. Unilateral contact forces can be determined from the requirement that $\eta_i>0$ implies sustained contact, i.e. $u_Z^T r_i=u_Z^Tv_i = u_Z^T dv_i/d\tau = 0$. The last constraint can be expanded using the time derivative of \eqref{eq:pontossebesseg}  as
\begin{align} 
u_Z^T\left(\frac{dv_c}{d\tau}+\frac{d\omega}{d\tau} \times (r_i-r_c)+\omega \times (\omega \times(r_i-r_c))\right)=0. \label{eq:rollc}
\end{align}
In the case of free fall, we have $\eta_i=0$ for all $i$ and the acceleration is determined by \eqref{eq:rollF} and \eqref{eq:rollM} as:
\begin{align}
\frac{d v_c}{d\tau}&=-gu_Z
\label{eq:acc1free}
\\
\frac{d \omega}{d\tau}&=-\theta^{-1}\left(  \omega \times \theta \omega \right)
\label{eq:acc2free}
\end{align}

 We will also investigate motion with one sustained contact, in which case $\frac{d v}{d\tau}$, $\frac{d \omega}{d\tau}$ and one unknown contact force $\eta_i$ are determined uniquely by the system of 7 scalar equation formed by \eqref{eq:rollF}, \eqref{eq:rollM} and \eqref{eq:rollc}:
\begin{align}
\frac{d v_c}{d\tau}&=u_Z\left(-g+\eta_i / m\right)
\label{eq:acc1contact}
\\
\frac{d \omega}{d\tau}&=\theta^{-1}\left(((r_i-r_c)\times \eta_i u_Z)-\omega \times \theta \omega \right)
\label{eq:acc2contact}
\\
\eta_i&=\left(g+u_Z^T(\omega \times R_{\times} \omega - R_{\times} \theta^{-1} (\omega \times
\Theta \omega))\right)/u_z^T[\theta^{-1}-R_{\times}\theta^{-1}R_{\times}]u_z
\end{align} 
 
 where $R_{\times}$ is the matrix representation of the cross product $(r_i-r_c)\times*$ (i.e. $R_{\times} x=(r_i-r_c) \times x$ for all $x \in \mathbb{R}^3$).
  Motion with two sustained contacts is determined similarly by using two constraints of type \eqref{eq:rollc}. More than 2 sustained contact implies that the whole disk lies on the floor.

\subsection{Impacts}

The object undergoes a single-point impact if one of the vertices hits the plane
\begin{align}
u_Z^Tr_i=0
\quad
u_Z^Tv_i<0
\end{align}
for some $i$ whereas all other vertices are separated.

Let $\zeta u_Z$ ($\zeta\in \mathbb{R}$) denote an instantaneous impulse the underlying plane exerts upon $\mathcal{B}$ in a single-point impact. The pre- and post-impact values of the velocity of the centre of mass, the angular velocity and the velocity of vertex $i$ will be distinguished by superscripts $^-$ and $^+$. The conservation of linear and angular momentum yield  
  
\begin{align}
m (v_c^{+}-v_c^{-})=\zeta u_Z \label{eq:utk1}\\
\theta (\omega^+-\omega^-) = r_i\times \zeta u_Z \label{eq:utk2}
\end{align}
We assume that all impacts have the same Newtonian coefficient of restitution $0\leq\gamma<1$, i.e. every impact satisfies
 \begin{align}
 u_Z^T v_i^+ = -\gamma u_Z^T v_i^- \label{eq:utk3}
 \end{align}
The unknowns $v_c^+$, $\omega^+$, and $\zeta$ are uniquely determined by the equations \eqref{eq:utk1}-\eqref{eq:utk3} as follows:
 \begin{align} 
\zeta&=\frac{-u_z^T(\gamma+1)(v_c^-+(\omega^- \times r_i))}{u_z^T\left[m^{-1} I - R_{\times} \theta^{-1} R_{\times} \right] u_z } \label{eq:F} \\
\omega^+&=\theta^{-1}((r_i-r) \times \zeta u_z)+\omega^-  \label{eq:omega+}\\
v_c^+&=m^{-1}\zeta u_z +v_c^- \label{eq:v+}
\end{align}
where $I$ stands for the 3 by 3 identity matrix.

Simultaneous impacts at multiple points occur if one vertex hits the ground while at least one other is also in contact with it. Our model for single impacts is applicable to this scenario is some cases. If vertex $i$ hits the ground while another vertex $j$ is in sustained contact (and all other vertices are separated), moreover \eqref{eq:F}-\eqref{eq:v+} predict post-impact velocities, which respect the unilateral contact constraint $u_Z^Tv_j^+\geq 0$, then we assume the single-impact model is valid. As we will see, this requirement is very often satisfied during the motion of an $n$-gon with $n>>1$. 

Simultaneous impacts not obeying this requirement are in general general highly unpredictable. Thus we do not attempt to specify a particular impact model. Instead, we declare that the dynamics of the disk is undecidable within our simple modelling framework if such a situation occurs.

\subsection{Linearization and general coordinates}
\label{sec:linearized}

 Without loss of generality, we may assume that the global and local frames coincide at the end of the motion. In addition, we assume that the disk becomes immobile after the singularity has been reached (importantly, no "yaw" motion with the local $z$ axis as axis of rotation is allowed).
 
We focus on the last phase of motion of the disk before reaching its singularity. Its rotations and angular velocities during this phase are close to 0. Rotations can be represented by a rotation vector $\phi\in\mathbb{R}^3$ where the direction of the vector represents the axis of the rotation and  $|\phi|$ corresponds to the angle of the rotation. In the case of small rotations, $\phi$ is equivalent of the rotation matrix 
\begin{align}
H_{l,g}=
\begin{bmatrix}
1 & -u_Z^T\phi & u_Y^T\phi\\
u_Z^T\phi & 1 & -u_X^T\phi\\
-u_Y^T\phi  & u_X^T\phi & 1
\end{bmatrix}
+O(|\phi|^2)
\label{eq:Hlin}
\end{align}
moreover $\phi$ is related to angular velocity as $\omega=\frac{d}{dt}\phi+O(|\phi|^2)$. 

If we neglect $O(|\phi|^2)$  terms in the kinematic equations, then\eqref{eq:pontoshelyzet} and \eqref{eq:Hlin} imply
\begin{align}
r_i^g=\colvec{3}
{x_i}
{y_i}
{u_Z^Tr_c + u_X^T\phi y_i  - u_Y^T\phi x_i} 
\label{eq:rig}
\end{align}
i.e. the distance of a vertex from $\mathcal{P}$ is expressed as a linear combination of the quantities $h:= u_Z^Tr_c$, $\phi_x:= u_X^T\phi$ and $\phi_y:= u_Y^T\phi$. This motivates our choice of the generalized coordinates  $q=(\phi_x,\phi_y,h)^T$ spanning a 3D subspace $\mathbb{C}$ of the 6D configuration space. An analogous 3D subspace $\mathbb{V}$ of the velocity space  is spanned by the generalized velocities $p=\frac{dq}{d\tau}=(\omega_x,\omega_y,v)^T$. Motion along the remaining 3 degrees of freedom (translational motion in $X$, $Y$ directions and rotation about $Z$) does not occur as long as contact forces are frictionless and as long as $O(|\omega|^2)$ terms in the equations of motion (see next subsection) are neglected.

Let us introduce the notation
\begin{align}
f_i:=\colvec{3}{y_i}{-x_i}{1}
\label{eq:fi}
\end{align}

Then the height of a vertex above $\mathcal{P}$ and its velocity in the global $Z$ direction can be expressed simply as 
\begin{align}
h_i=f_i^T q\;\;\; v_{i}=f_i^Tp. \label{eq:hi,vi}
\end{align}

\subsection{Linearized equations of motion}

The linearized equations of motion in the generalized coordinates are obtained by elimination of $O(|\phi|^2)$ and $O(|\omega|^2)$ terms in the equations\eqref{eq:acc1free}, \eqref{eq:acc2free}, \eqref{eq:acc1contact}, \eqref{eq:acc2contact}, \eqref{eq:omega+}, and \eqref{eq:v+}. For free fall we obtain the equation
\begin{align}
\frac{dp}{d\tau}=-gu_3
\label{eq:paccfree}
\end{align}   
where $u_3=[0,0,1]^T$. 
Note that this means that the acceleration of every vertex during free fall is constant $-g$ by \eqref{eq:rig},\eqref{eq:paccfree}. 

During motion with a single point contact, we similarly obtain 
\begin{align}
 \frac{dp}{d\tau}=\Theta^{-1}\left[\eta f_i-mgu_3\right]:=a_{contact}
 \label{eq:pacccontact}
\end{align}  
with
\begin{align}
\Theta=
  m\begin{bmatrix}
    \rho^2 & 0 & 0 \\
    0 & \rho^2 & 0 \\
    0 & 0 & 1
  \end{bmatrix}
  \label{eq:Theta}
\\
\eta=\frac{mg}{1+\rho^{-2}} 
\end{align}
During this motion, the accelerations of the vertices are unequal, but all of them are negative except for point $i$ having 0 acceleration.

Finally, an impact ar vertex $i$ corresponds to the linear map 
\begin{align}
p^+=U_ip^-\label{eq:impactmap}
\end{align} 
where 
\begin{align}
U_i:=I+\frac{-(1+\gamma)}{f_i^T \Theta^{-1}f_i}\left[\Theta^{-1}f_i f_i^T \right]\label{eq:impactequ}
\end{align} 

\subsection{Invariance properties of the linearized  dynamics} \label{sec:invar} 
Our model of the imperfect disk possesses $n$-fold rotational symmetry. This means that the motion is invariant with respect to certain rotations. Let $q_{q_0,p_0,t_0}(t)$ and $p_{q_0,p_0,t_0}(t)$ denote the position and velocity of a disk at time $t$ in the case of initial conditions $q=q_0$, $p=p_0$ at time $t=t_0$. Then, 
\begin{align}
q_{Rq_0,Rp_0,t_0}(t)&=Rq_{q_0,p_0,t_0}(t)
\label{eq:Rq}\\
p_{Rq_0,Rp_0,t_0}(t)&=Rp_{q_0,p_0,t_0}(t)
\label{eq:Rp}
\end{align}
for any $t>t_0$ where
\begin{align}
R=
  \begin{bmatrix}
    cos(2\pi/n) & -sin(2\pi/n) & 0 \\
    sin(2\pi/n) & cos(2\pi/n) & 0 \\
    0 & 0 & 1
  \end{bmatrix}
  \label{eq:R}
\end{align}
In addition, the linearized equations of motion also admit the following scaling invariance for any $\beta\in\mathbb{R}$:
\begin{align}
q_{\beta^2q_0,\beta p_0,t_0}(t_0+\beta t)&=
\beta^2q_{q_0,p_0,t_0}(t_0+t)
\label{eq:scaleq}
\\
p_{\beta^2q_0,\beta p_0,t_0}(t_0+\beta t)&=
\beta p_{q_0,p_0,t_0}(t_0+t)
\label{eq:scalep}
\end{align}
We will exploit these invariance relations in order to define low-dimensional reduced Poincaré maps, whose invariant points correspond to interesting forms of motion of the disk.

\section{Analysis of precession-free motion}

As we have seen, the motion of Euler's disk on the short run appears to be similar to the precession-free circular motion of a dissipation-free rolling disk whose center of mass remains immobile. On the long run, the disk tends to drift along a one-parameter family of such solutions as energy is dissipated. This motion corresponds to a $\phi$ rotation vector, which rotates continuously about the global $Z$ axis, while $|\phi|$ is decreasing monotonically. Our model of the imperfect disk is not capable of dissipation-free rolling motion because its polygonal shape creates impacts. Nevertheless,we will define a certain form of motion reminiscent of precession-free rolling of a round disk. First, we discuss the case of inelastic collisions ($\gamma=0$) in detail and then we review the consequences of an impact model with $\gamma>0$ in the last subsection.

\subsection{Self-similar motion}

An ideally inelastic collision means that every vertex stays on the ground after it has undergone an impact. This kinematic constraint implies that only one of its neighbours may hit the ground next. Assume that an impact at vertex $i$ is followed by another one at $i+1$. It is easy to show that for $n\geq 4$, $v_i$ will always become positive at the second impact, and thus the only possible order of impact locations is 0,1,...,$n-1$,0,1,.... (In contrast, we note without proof that for $n=3$ and sufficiently small $\rho$, a sequence $...,i,i+1,i,...$ is also possible.) This motion appears to be "precession-free" if the values of $|\phi|$ evaluated at the times of impacts form a monotonically decreasing sequence. The invariance properties of Sec. \ref{sec:invar} suggest the following definition characterizing motion similar to the precession-free rolling of a round disk:
\begin{definition}[Self-similar motion]
A set of initial conditions $q_0$, $p_0$, $t_0$ satisfying 
\begin{align}
f_0^Tq_0=0 \label{eq:ftq0}\\
f_0^Tp_0<0
\end{align} 
(pre-impact state at vertex 0) as well as the motion initiated by such a set are called self-similar if 
\begin{itemize}
\item the motion starts with an immediate impact at vertex 0 followed by impact-free motion until vertex 1 hits the ground at time $t_1$
\item  immediately before the second impact of the system, the generalized velocity and coordinates are
\begin{align}
p_1=\beta R p_0\label{eq:selfsimilar1}\\
q_1=\beta^2 R q_0\label{eq:selfsimilar2}
\end{align} with some scalar $0<\beta<1$.
\end{itemize}
\label{def:selfsimilar}
\end{definition}
 The conditions \eqref{eq:selfsimilar1}-\eqref{eq:selfsimilar2} and the invariance relations \eqref{eq:Rq},\eqref{eq:Rp}, \eqref{eq:scalep},\eqref{eq:scaleq} imply that "self-similar motion" then continues with rotated and downscaled copies of the motion during the time interval $(t_0,t_1)$.  More specifically, if $t_i$ denotes the time of the $(i+1)^{th}$ impact, $q_i$ and $p_i$ are the corresponding pre-impact values of state variables, and $\tau_i=t_{i+1}-t_i$ then 
the vertices hit the ground in the order 0,1,...,$n-1$,0,1,...; each vertex remains in contact with the ground until the impact at the next vertex occurs; the durations decrease exponentially as
\begin{align}
\tau_i&=\beta^i\tau_0
\label{eq:tauscaling}
\end{align}
whereas the vectors of state variables decrease exponentially and rotate around as follows:
\begin{align}
p_i&=\beta^iR^ip_0
\label{eq:pscaling}\\
q_i&=\beta^{2i}R^iq_0
\label{eq:qscaling}
\end{align}
Note that the sequence of $q_i$ vectors corresponds to an exponentially decreasing sequence of $|\phi|$ values if $0<\beta<1$, which makes self-similar motion similar to precession-free rolling of a round disk.

In order to find self-similar initial conditions, let us expand \eqref{eq:selfsimilar1} and \eqref{eq:selfsimilar2} as:

\begin{align}
(U_0p_0+ a_{contact}\tau_0) = R\beta p_0\label{eq:pscaling_exp}\\
(q_0+U_0p_0 \tau_0+a_{contact}\tau_0^2/2)= R\beta^2 q_0\label{eq:qscaling_exp}
\end{align}

If the matrix $R^{-1}U_0$ has a real eigenvalue $\lambda$, then these equations have a corresponding trivial solution: $\tau_0=0$; $q_0=0$; $\beta=\lambda$ and $p_0=$ the eigenvector corresponding to $\lambda$. Eigenvalue analysis of $R^{-1}U_0$ reveals one trivial solution with $\beta>0$ (if $n=3$) or two trivial solutions with $\beta>0$ (if $n>3$) for sufficiently small values of $\rho$ (star markers in Fig. \ref{fig:plasticbeta}).

In contrast, if $\beta$ is not a real eigenvalue of $R^{-1}U_0$, then we can express $q_0$ and $p_0$ explicitly from these equations (in terms of the unknown scalars $\tau_0$ and $\beta$) as 
\begin{align}
p_0=(\beta I -R^{-1}U_0)^{-1} R^{-1} \ a_{contact}\tau_0
\label{eq:p0exp}\\
q_0=(\beta^2 I -R^{-1})^{-1} R^{-1}(U_0 p_0\tau_0+a_{contact}\tau_0^2/2).
\label{eq:q0exp}
\end{align}
where $I$ is an identity matrix.
These two expressions can be inserted into the condition \eqref{eq:ftq0}, yielding an equation of the form
$$
\tau_0^2 \pi(\beta)=0
$$
where $\pi$ is a polynomial of degree 9. This is satisfied on the one hand if $\tau_0=0$, which again recovers the trivial solution. On the other hand $\tau_0\neq 0$ means that we must have $\pi(\beta)=0$. The roots of $\pi$ were found numerically (circle markers in Figure \ref{fig:plasticbeta} for various values of $n$ and the radius of gyration $\rho$.  We have found at least 1 positive root for all combinations of model parameters and for all values of $n$.
Each root for $\beta$ corresponds to a one-parameter family of initial conditions, parametrized by $\tau_0$ as given by \eqref{eq:p0exp},\eqref{eq:q0exp}. Each family of solutions is generated from one solution by the scaling invarance\eqref{eq:scaleq}, \eqref{eq:scalep}, hence we can consider infividual solution within each family as identical.

\subsection{Feasibility of self-similar motion}
Solutions found by this numerical procedure may be infeasible. Feasibility means that 
\begin{enumerate}
\item the initial (pre-impact) velocity of vertex 0 must point downwards, i.e. $f_0^Tp_0<0$
\item the duration of continuous motion between the first two impacts must be positive, i.e. $\tau_0>0$. 
\item none of the vertices penetrates into the ground during  the time interval $t_0<t<t_1$. It is easy to show that checking non-penetration at time $t_0$ is sufficient.
\item the $\beta$ value associated with the motion is positive.
\end{enumerate}

The trivial solutions are either infeasible or marginally feasible
because $\tau_0=0$. 
The significance of these solutions lies in that gravity has no time to act if $\tau=0$. As we will see later, there may exist feasible solutions asymptotically converging to the trivial solutions, which do not rely on the effect of the gravitational force.

\begin{figure}[h]
\begin{center}
\begin{subfigure}[t]{0.5\textwidth}
\includegraphics[width=8cm]{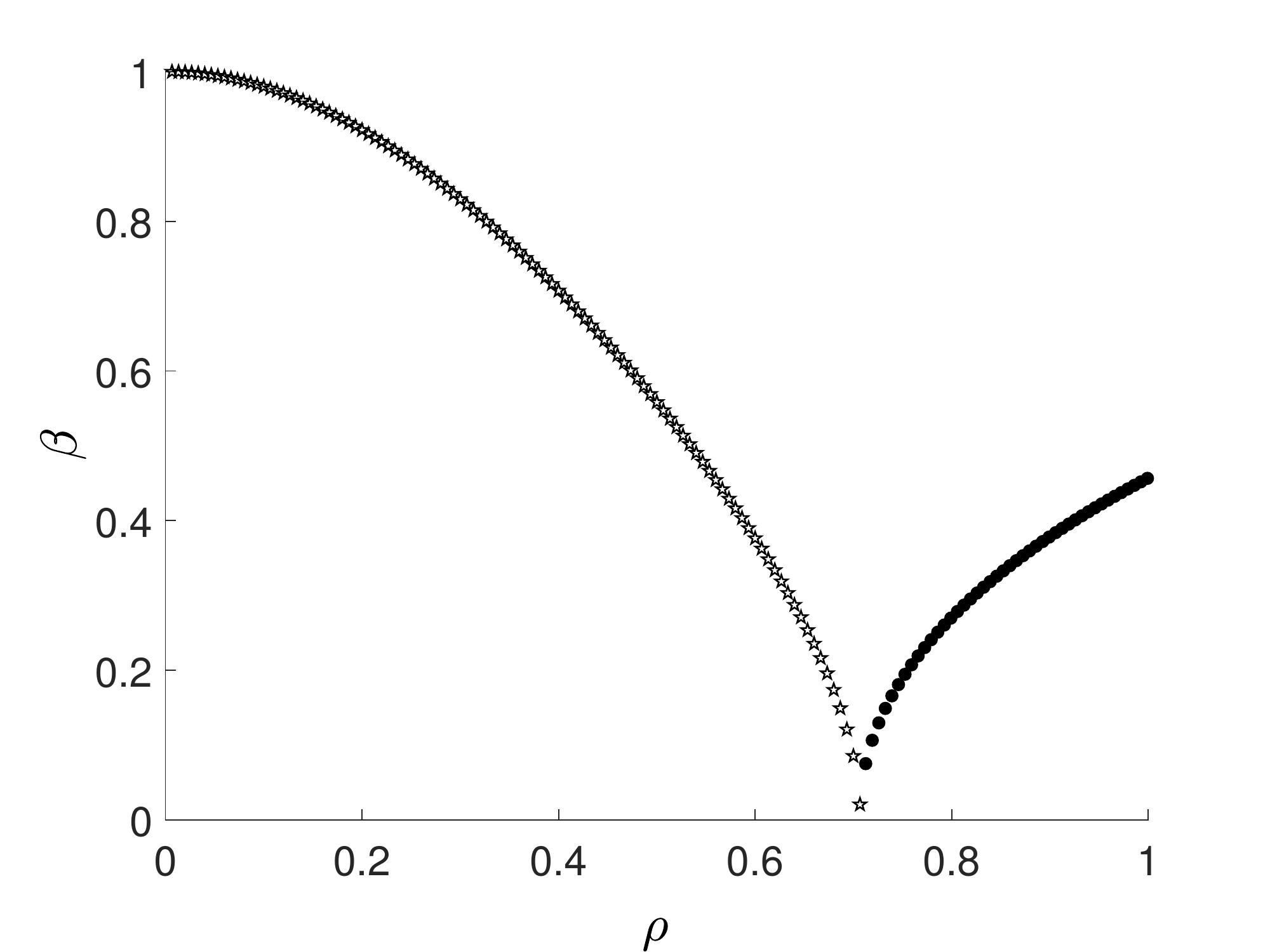}
\caption{$n=3$}
\end{subfigure}%
\begin{subfigure}[t]{0.5\textwidth}
\includegraphics[width=8cm]{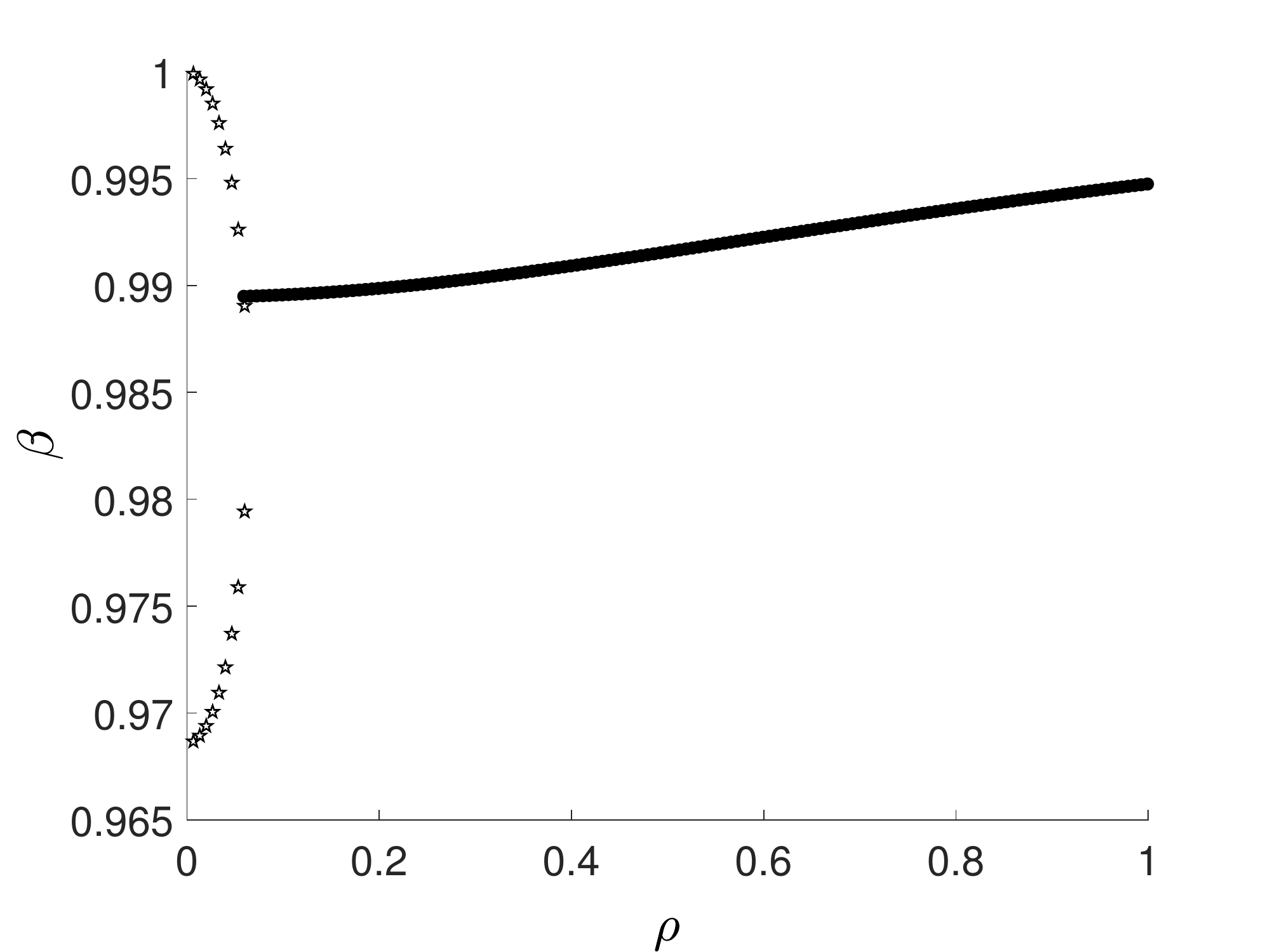}
\caption{$n=25$}
\end{subfigure}
\caption{Values of $\beta$ depending on the physical properties of the poligonal model}\label{fig:plasticbeta}
\end{center}
\end{figure}

\begin{figure}[h]
\begin{center}
\includegraphics[height=9cm]{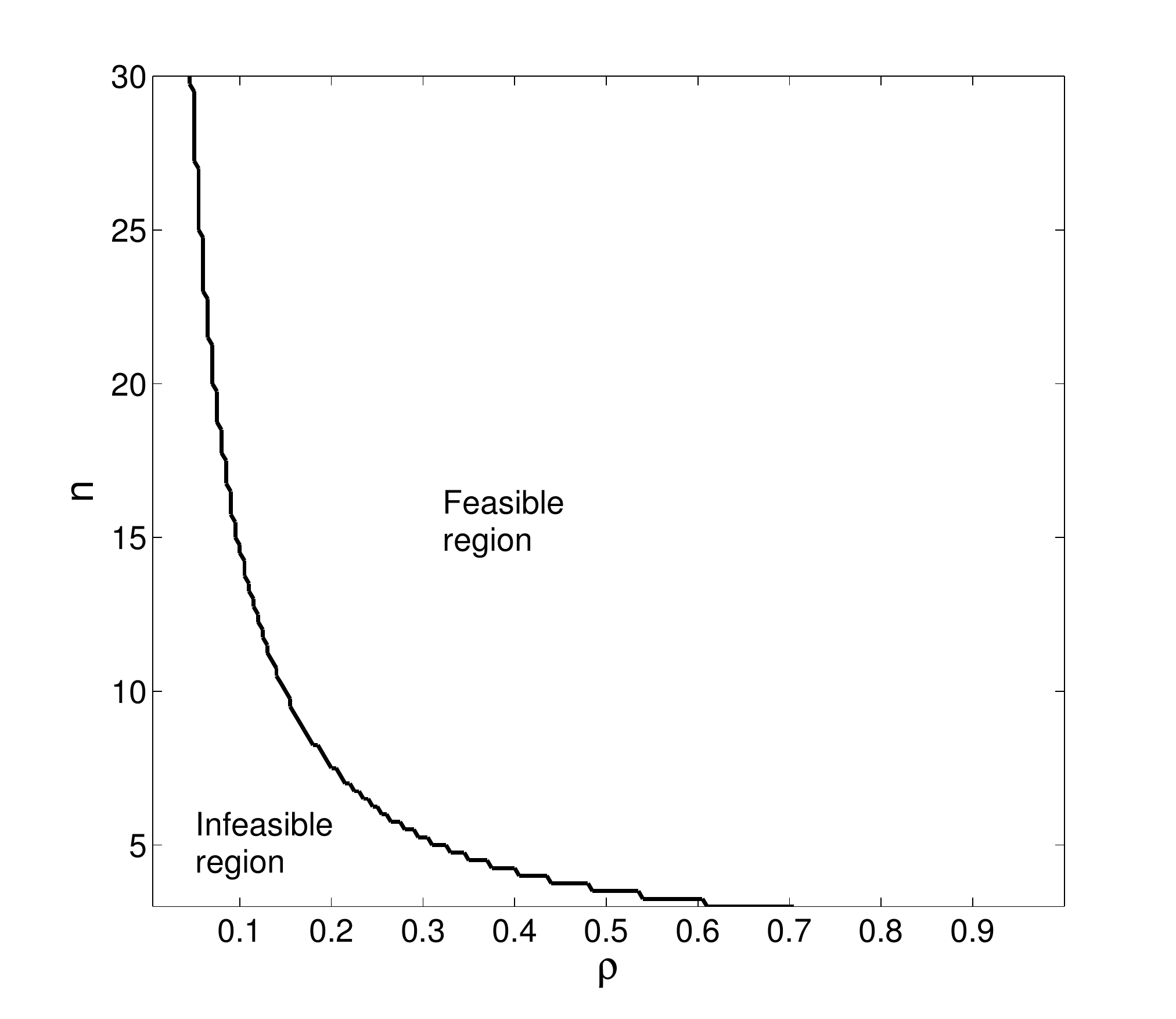}
\caption{Border of feasible and infeasible regions for self similar motion with $\gamma=0$. Notice that homogenous disk corresponds to $\rho\approx 1/2$, in which case the non-trivial self-similar solution is feasible for all $n>3$}
\label{fig:c0_feas}
\end{center}
\end{figure}

We checked the feasibility of the nontrivial solutions numerically. and found that the nontrivial solutions are feasible for large enough values of $\rho$ whenever $n>3$ (filled circles in Fig. \ref{fig:plasticbeta}, see also Fig. \ref{fig:c0_feas}). The most important conclusion for the spinning disk problem is that there is a unique feasible, self-similar motion for a polygonal approximation of a thin, round disk ($n>>1$) with homogenous density ($\rho \approx 0.5$). 

\subsection{Stability analysis of self-similar motion}

Whether or not the nontrivial self-similar motion found above is realized by the system depends largely on the stability of that type of motion. We will test the asymptotic stability of self-similar mktion by investigating a smooth discrete map induced by the dynamics (similar to the Poincaré maps of continuous dynamical systems), whose invariant points are in one-to-one correspondence with self-similar initial conditions.

From now on, we assume that the disk moves in such a way that the vertices hit the ground in the order $0,1,2,...,n-1,0,1,...$ (which is definitely true for small perturbations of self-similar motion and for a limited amount of time). Nevertheless we do not require that the motion is self-similar.

As before, $p_0$, $q_0$ denote values of the state variables immediately before an impact at vertex 0, and $p_i$, $q_i$ ($i=1,2,...$) denote the subsequent pre-impact values of the state variables. Each of these impacts then occurs at vertex $i\mod(n)$. Let $C_i$ denote the maps that transform vectors composed of these state variables into the next one as follows:
$$
\begin{bmatrix}
p_{i+1}\\q_{i+1}
\end{bmatrix} =C_i
\left(
\begin{bmatrix}
p_i\\q_i
\end{bmatrix}
\right)
$$
Then, $C_i$ represents the effect of an impact at vertex $i\mod(n)$ and the subsequent continuous motion.

 We also introduce a collection of maps $T_i$ ($i=1,2,...$), which rotate and normalize vectors of stacked state variables as:
\begin{align}
\begin{bmatrix}
R^{-i}p/|p|
\\
R^{-i}q/|p|^2
\end{bmatrix}
=
T_i\left(
\begin{bmatrix}
p
\\
q
\end{bmatrix}
\right)
\end{align}
as well as the transformed state variables
$$
\begin{bmatrix}
\bar{p}_{i}\\\bar{q}_{i}
\end{bmatrix} 
=
T_i
\left(
\begin{bmatrix}
p_i\\q_i
\end{bmatrix}
\right)
$$
and the map
\begin{align}
\bar{C}(*)=T_1(C_0(*))
\label{eq:barC}
\end{align}
Then, the scaling and rotational invariance of the dynamics (Sec. \ref{sec:invar}) imply
\begin{align}
\begin{bmatrix}
\bar{p}_{i+1}\\\bar{q}_{i+1}
\end{bmatrix} 
=\bar{C}
\left(
\begin{bmatrix}
\bar{p}_i\\\bar{q}_i
\end{bmatrix}
\right)
\label{eq:Cbariteration}
\end{align}
i.e. the dynamics of the system can be understood by studying the iteration defined by the map $\bar{C}$. 

The map $\bar{C}$ can be expressed in closed from as 
\begin{align}
\begin{pmatrix}
p\\q 
\end{pmatrix}
\mapsto 
\begin{pmatrix}
R^{-1}(U_0p+ a_{contact}\tau)/ \left|R^{-1}(U_0p+ a_{contact}\tau\right|
\\
 R^{-1}(q+U_0p \tau+a_{contact}\tau^2/2) / \left|
 R^{-1}(U_0p+ a_{contact}\tau)\right|^{2}
\end{pmatrix}
\label{eq:barCexpression01}
\end{align}

where 
\begin{align}
\tau=\frac{-f_1^TU_0p+\sqrt{(-f_1^TU_0p^2-2(f_1^Ta_{contact})(f_1^Tq))}}{f_1^Ta_{contact}}
\label{eq:barCexpression02}
\end{align}
(see Appendix \ref{sec.barC} for details). This map is formally defined over the six dimensional space $\mathbb{V}\times\mathbb{C}$. Every point is mapped into the 2D subspace defined by 
\begin{align}
f_0^Tq=0,|p|=1,f_{n-1}^Tq=0,f_{n-1}^Tp=0\label{eq:outputspace}
\end{align}
Each one-parameter family of self-similar initial conditions has one member with $|p|=1$, and these members are in one-to-one correspondance with the invariant points of $\bar{C}$. The asymptotic stability of these invariant points can be examined by eigenvalue analysis of the Jacobian of $\bar{C}$.
The existence of an eigenvalue with absolute value greater than 1 implies asymptotic instability and if all eigenvalues have absolute values below 1, then the invariant point is asymptotically stable.  

In our case, the Jacobian always has 0 as a four-fold eigenvalue due to \eqref{eq:outputspace}. In order to determine the remaining two eigenvalues, we
expressed the Jacobian in closed form by differentiating \eqref{eq:barCexpression01}. The eigenvalues of the Jacobian were found numerically at the (numerically obtained) invariant points of $\bar{C}$.

The analysis outlined above has been completed for various values of $n$ and $\rho$ and we found that a feasible self-similar motion is always asymptotically stable. This is a local result, which does not imply that the motion of the disk must globally converge to self-similar motion from arbitrary initial condition. Nevertheless it suggests that the assumption of precession-free motion (analogous to self-similarity in our model) often used for the analysis of Euler's disk is plausible.

\subsection{Energy dissipation during self-similar motion}\label{sec:energydissipation}
Now we are ready to investigate energy dissipation due to impacts as the disk approaches its singularity. We now adopt the hypothesis that the polygonal disk undergoes self-similar motion as explained in the previous sections. The scaling law \eqref{eq:tauscaling} leads to a finite-time singularity at time
$$
t_{f}:=t_0+\sum_{i=0}^{\infty}\tau_i=t_0+(1-\beta)^{-1}\tau_0
$$
which can also be expressed as
\begin{align}
t_{f}-t_i=\sum_{j=i}^{\infty}\tau_j=\beta^i(1-\beta)^{-1}\tau_0
\label{eq:tdiffscaling}
\end{align}
for arbitrary $i$.

At the same time, the potential energy of the disk can be expressed as a homogenous linear function of $q$ whereas the kinetic energy is a homogenous quadratic function of $p$. The rotational invariance and the scaling laws \eqref{eq:pscaling}, \eqref{eq:qscaling} imply that each one of the potential energy, the kinetic energy as well as their sum in the pre-impact states at time $t_i$ form exponentially decreasing sequences. In particular, if $E_i$ is the pre-impact value of the total mechanical energy at $t=t_i$, then
\begin{align}
E_i=\beta^{2i}E_0
\label{eq:Escaling}
\end{align}

The equations \eqref{eq:tdiffscaling} and \eqref{eq:Escaling} yield the relation
\begin{align}
E_i=E_0\tau_0^{-2}(1-\beta)^2(t_f-t_i)^2
\label{eq:Ei}
\end{align}

 The total energy of the system is a piecewise constant function (with jumps at impact times), hence it cannot be expressed in the form \eqref{eq:scalinglaw}. Nevertheless by using \eqref{eq:tdiffscaling} and \eqref{eq:Ei}, it is easy to show that the bounds
\begin{align}
a_2(t_f-t)^c<E(t)<a_1(t_f-t)^c
\label{eq:scalinglaw2}
\end{align}
 are satisfied for all $t<t_f$ with 
\begin{align}
c&=2\\
a_1&=E_0\tau_0^{-2}(1-\beta)^2\\
a_2&=\beta^2 E_0\tau_0^{-2}(1-\beta)^2
\end{align}
We conclude that the exponent of energy dissipation due to impacts is 2. This is higher than the exponents associated with other dissipation mechanisms, hence we may draw the conclusion that impacts due to shape imperfection of the disk do not become dominant during the last phase of the motion. 

\subsection{Self-similar motion under partially elastic impacts} 

The assumption of inelastic impacts in the previous subsections was merely an a priori modelling assumption. It is an interesting question if the conclusions of the analysis with respect to the existence, and the stability of self-similar motion and the scaling of energy dissipation are sensitive to such modelling assumptions or not. 

In the case of partially elastic collisions ($0<\gamma<1$), Definition \ref{def:selfsimilar} remains applicable, however impacts are typically followed by free-flight rather than by sustained contact. 

\begin{figure}[h]
\begin{center}
\begin{subfigure}[t]{0.5\textwidth}
\includegraphics[height=5cm]{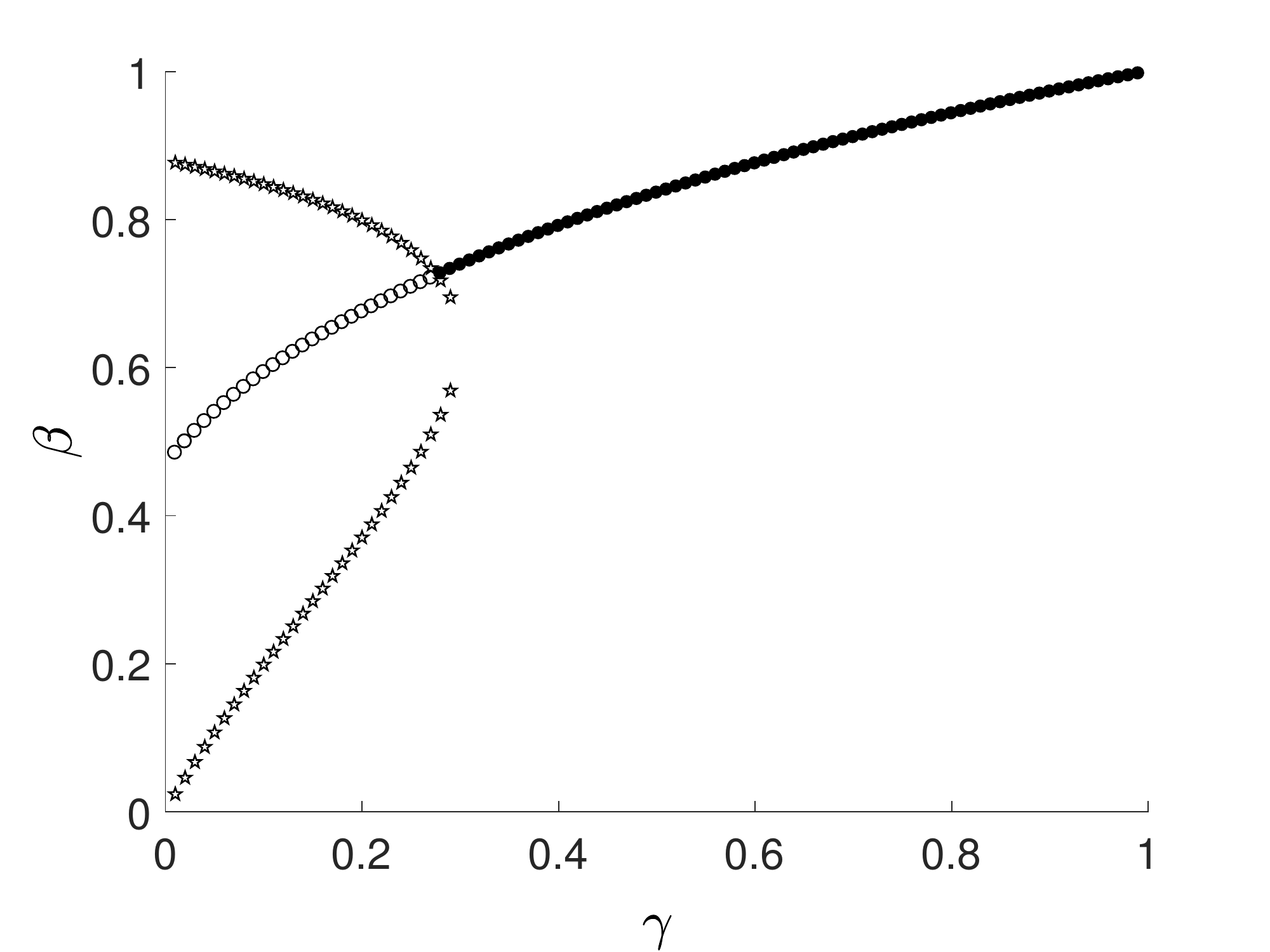}
\caption{$n=3$ and $\rho=0.25$}
\end{subfigure}%
\begin{subfigure}[t]{0.5\textwidth}
\includegraphics[height=5cm]{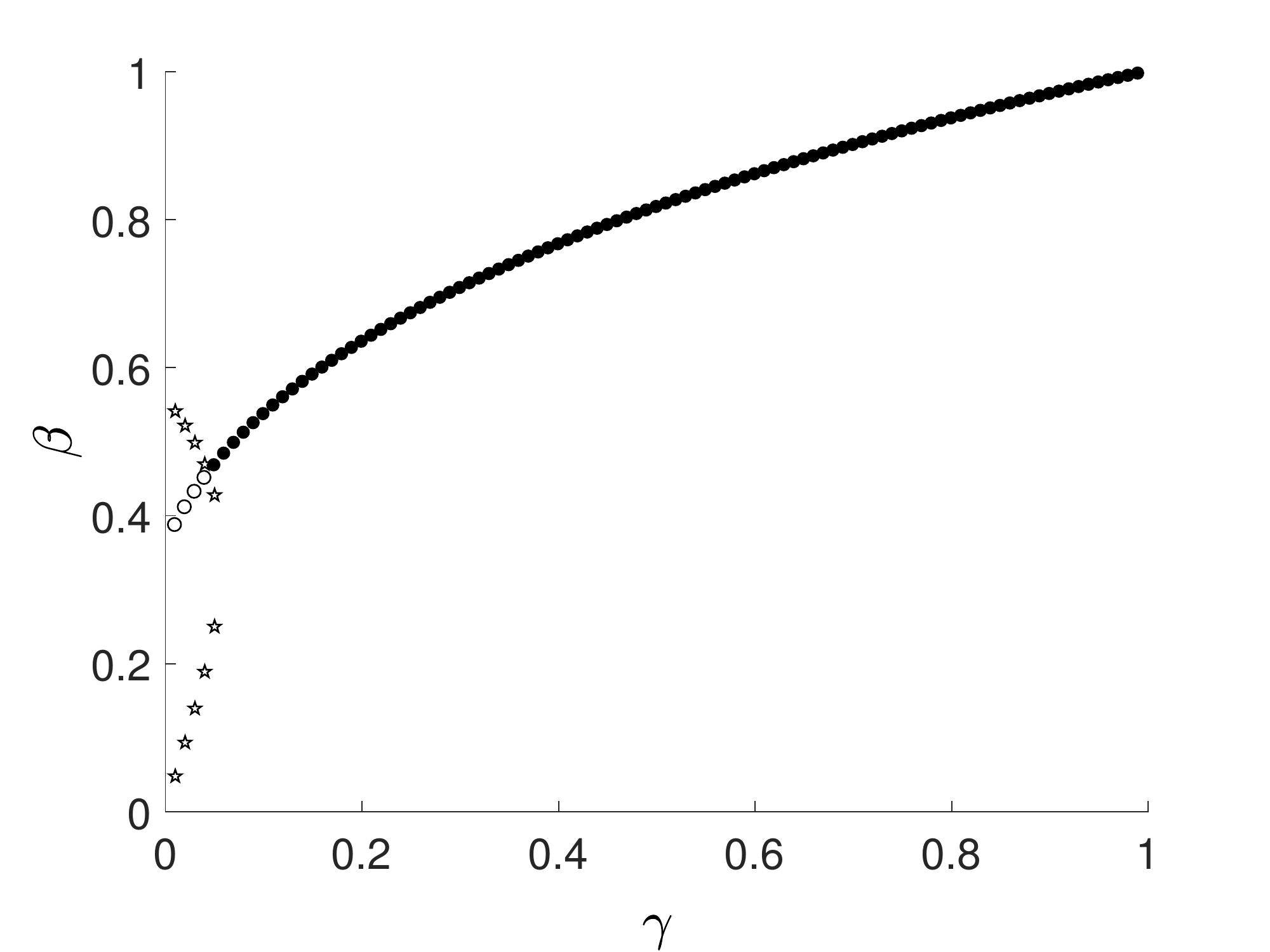}
\caption{$n=3$ and $\rho=0.5$}
\end{subfigure}
\\
\begin{subfigure}[t]{0.5\textwidth}
\includegraphics[height=5cm]{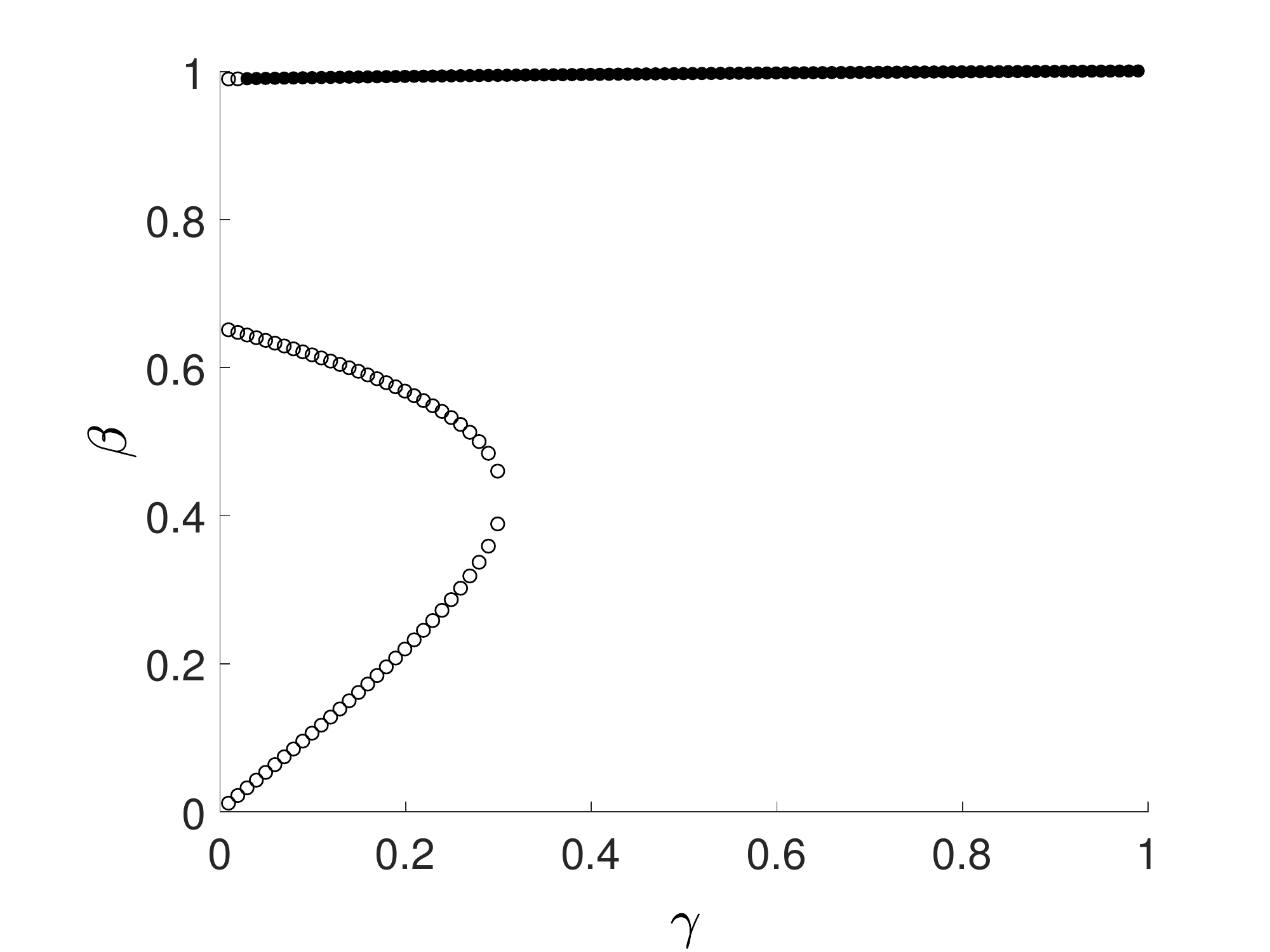}
\caption{$n=25$ and $\rho=0.25$}
\end{subfigure}%
\begin{subfigure}[t]{0.5\textwidth}
\includegraphics[height=5cm]{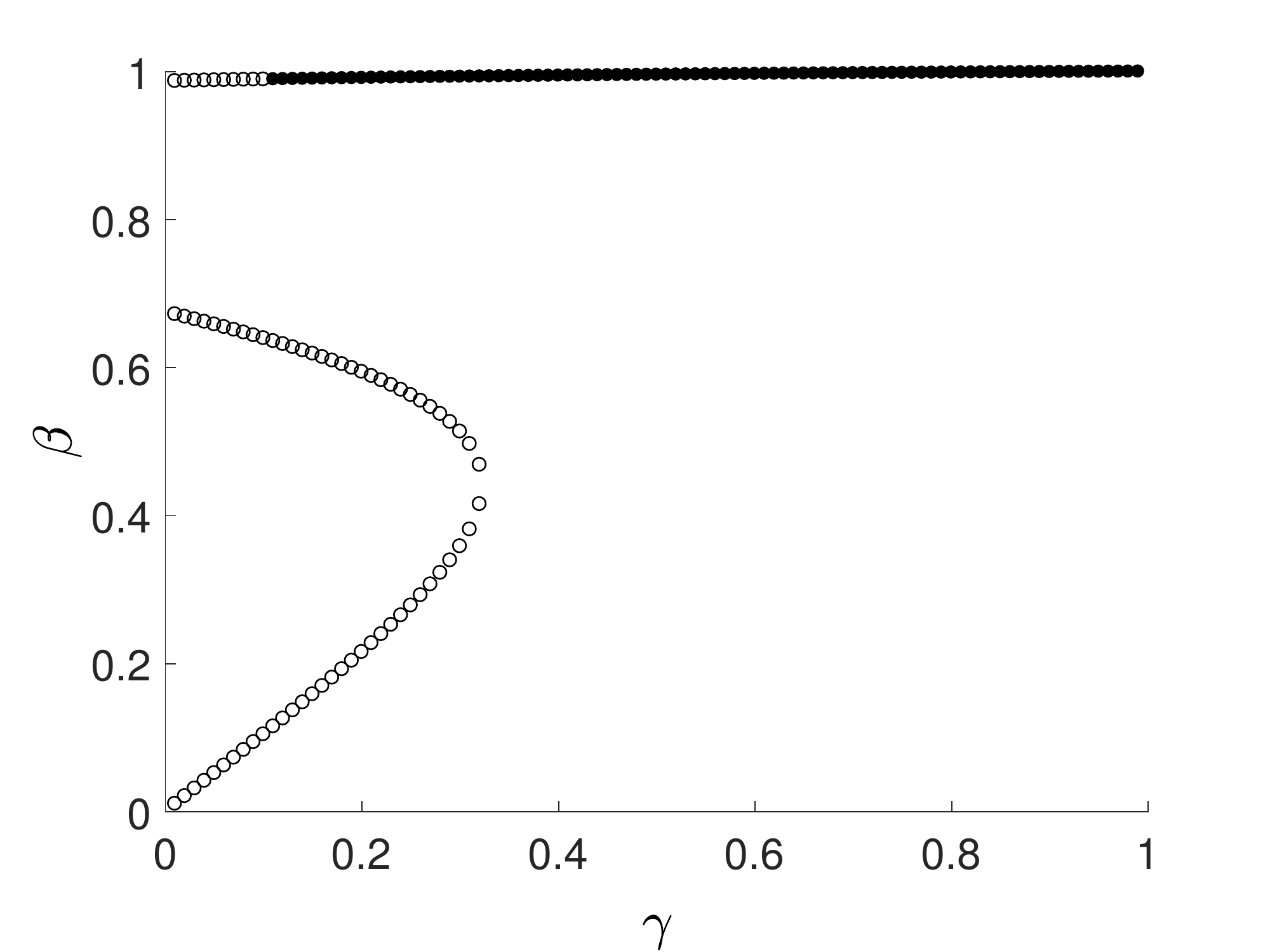}
\caption{$n=25$ and $\rho=0.5$}
\end{subfigure}
\caption{Trival (stars), and feasible (full circle) or infeasible (empty circle) nontrivial solutions to the equations of self-similar motion.}
\label{fig:feas}
\end{center}
\end{figure}

\begin{figure}[h]
\begin{center}
\includegraphics[height=8cm]{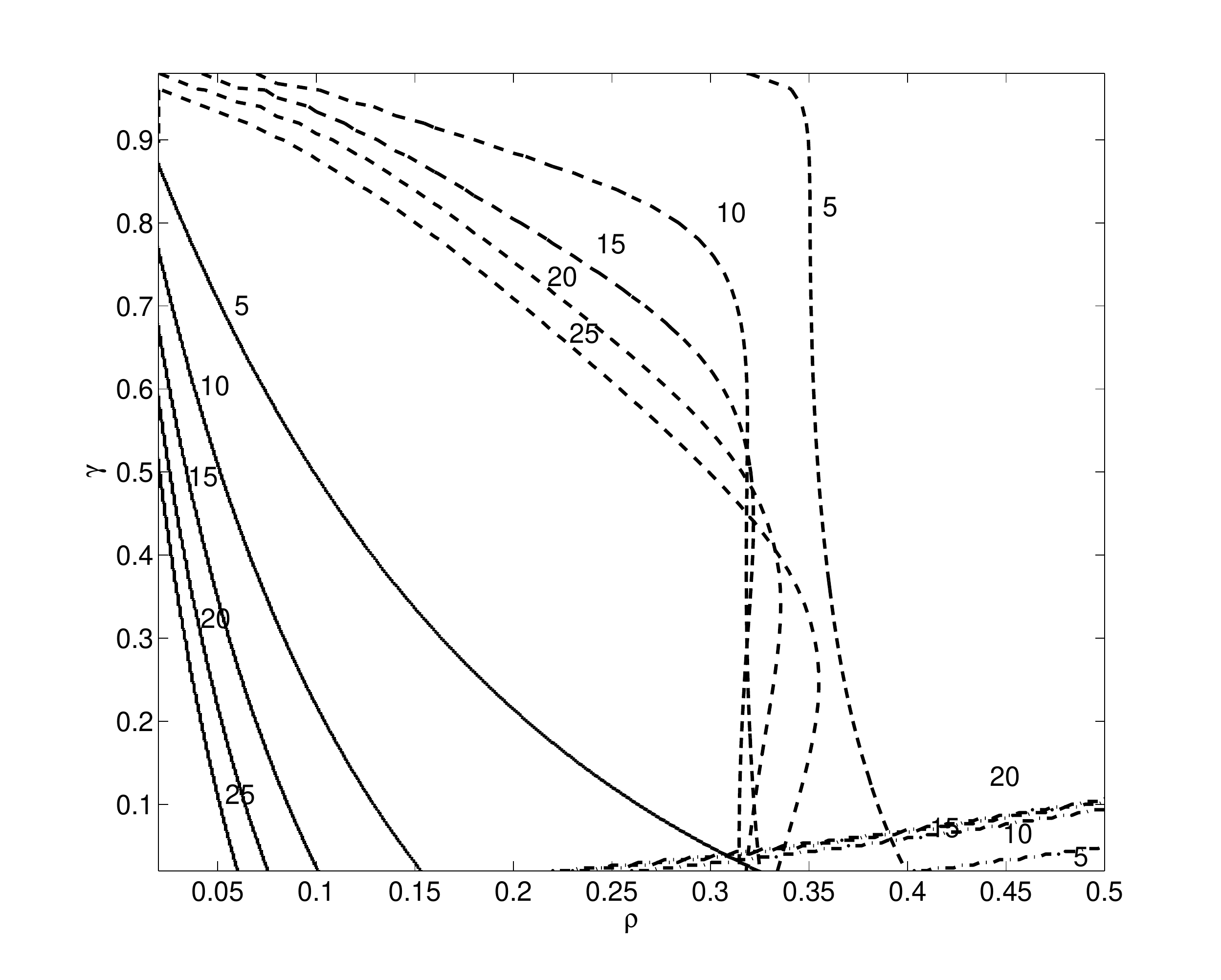}
\caption{Feasibility and stability of self-similar motion. At least one nontrivial solution exists for all parameter values. One feasible nontrivial solution exists above the dash-dotted lines (each curve corresponds to a specific value of $n$), and the feasible solution is stable for parameter values on the left of the dashed lines. Trivial solutions exist for parameter values below the solid lines.}
\label{fig:boundaries}
\end{center}
\end{figure}

We can perform the same analysis as in the preceding poarts of Sec. 3 with the only difference being that the equations \eqref{eq:pscaling_exp}, \eqref{eq:qscaling_exp} are now replaced by 
\begin{align}
\beta p_0=R^{-1}(U_0p_0-u_3 g\tau_0)\label{eq:similarpfly}\\
\beta^2 q_0=R^{-1}(q_0+p_0 \tau_0 -u_3 g \tau_0^2/2).\label{eq:similarqfly}
\end{align}
As before, we find marginally feasible trivial solutions (for small values of $\rho$ and $\gamma$) as well as one-parameter families of feasible or infeasible  nontrivial solutions parametrized by $\tau_0$. The results of the analysis are illustrated by Fig. \ref{fig:feas}, where $\beta$ values are depicted as functions of $\gamma$ for $\rho=0.5$ and for various values of $n$. We also summarized the results in Fig. \ref{fig:boundaries} for all combinations of $\beta$, $\gamma$ and $n$ values. In some cases multiple self-similar motions have been found, with different $\beta$ values. 
%
%
%
%





To examine the attractivity of self-similar motion, we again define a map $\bar{C}$ satisfying \eqref{eq:Cbariteration}. The explicit formula of the map is  similar to \eqref{eq:barCexpression01}, \eqref{eq:barCexpression02} except that $a_{contact}$ is replaced by $gu_3$. This map projects the six dimensional space $\mathbb{V}\times\mathbb{C}$ onto a four dimensional subspace given by
\begin{align}
f_0^Tq=0,|p|=1\label{eq:outputspace02}
\end{align}
Accordingly, the Jacobian of the map has now 0 as a trivial, two-fold eigenvalue. Numerical investigation of the nontrivial eigenvalues of the Jacobian revealed, that the stability of the motion depends on the parameters $n$, $\gamma$ and $\rho$. Most importantly, in the case of $\rho=0.5$ (the case of a homogenous disk) self-similar motion is unstable for all values of $n$ and $\gamma$. Hence the currently investigated model predicts that micro-impacts destabilize the precession-free motion of a spinning disk. Together with the results of the previous section, our findings highlight the sensitivity of the dynamics to fine details of the contact model (like the coefficient of restitution). Our numerical results concerning the existence, feasibility, and stability of self-similar motion are summarized in Fig. \ref{fig:boundaries}.

\begin{figure}[h]
\label{fig:edissip}
\centering
    \begin{subfigure}[t]{0.5\textwidth}
        \centering
        \includegraphics[width=65mm]{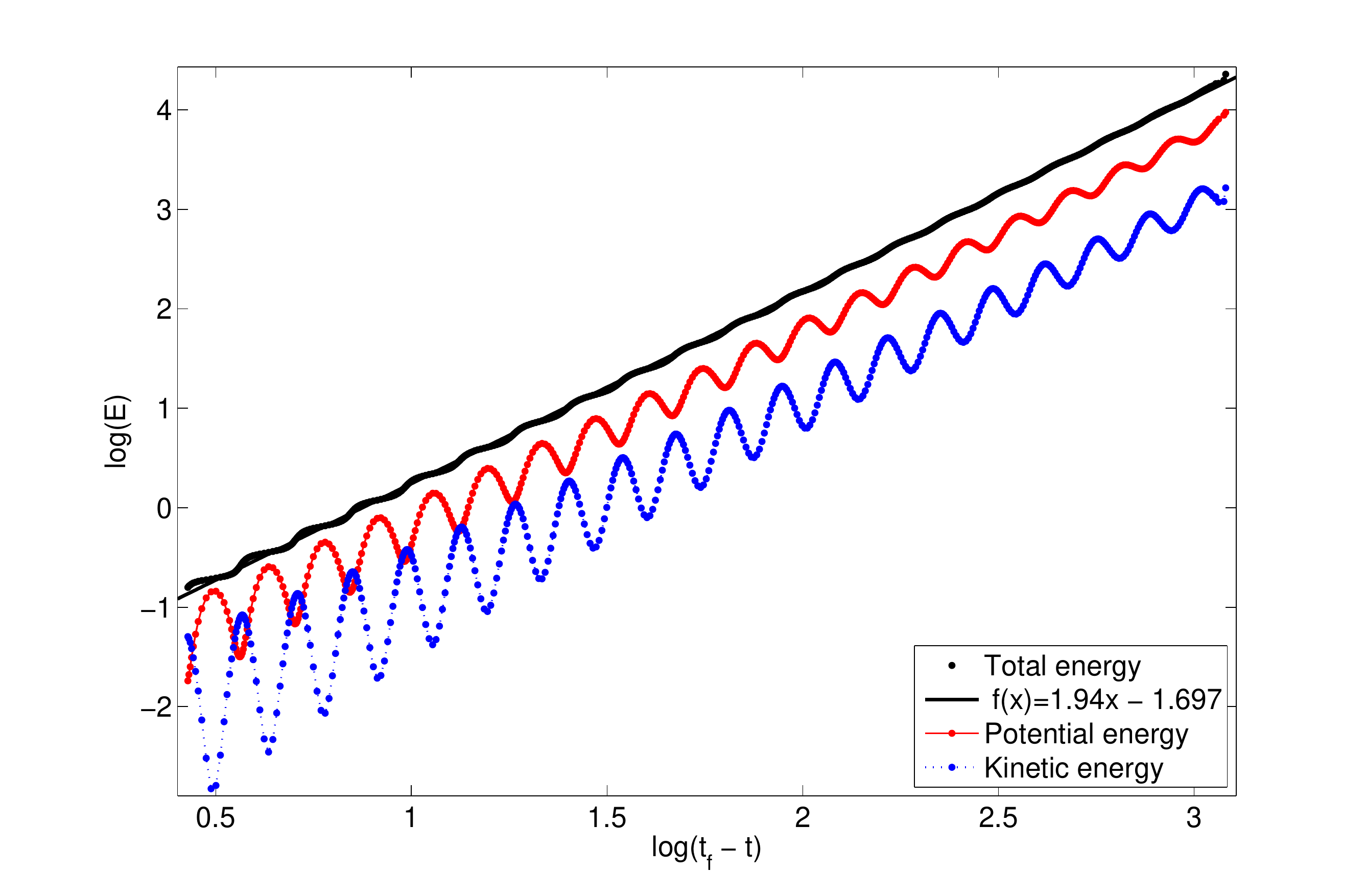}
        \caption{$n=20$ $\gamma=0.6$}
            \end{subfigure}%
    ~ 
    \begin{subfigure}[t]{0.5\textwidth}
        \centering
        \includegraphics[width=65mm]{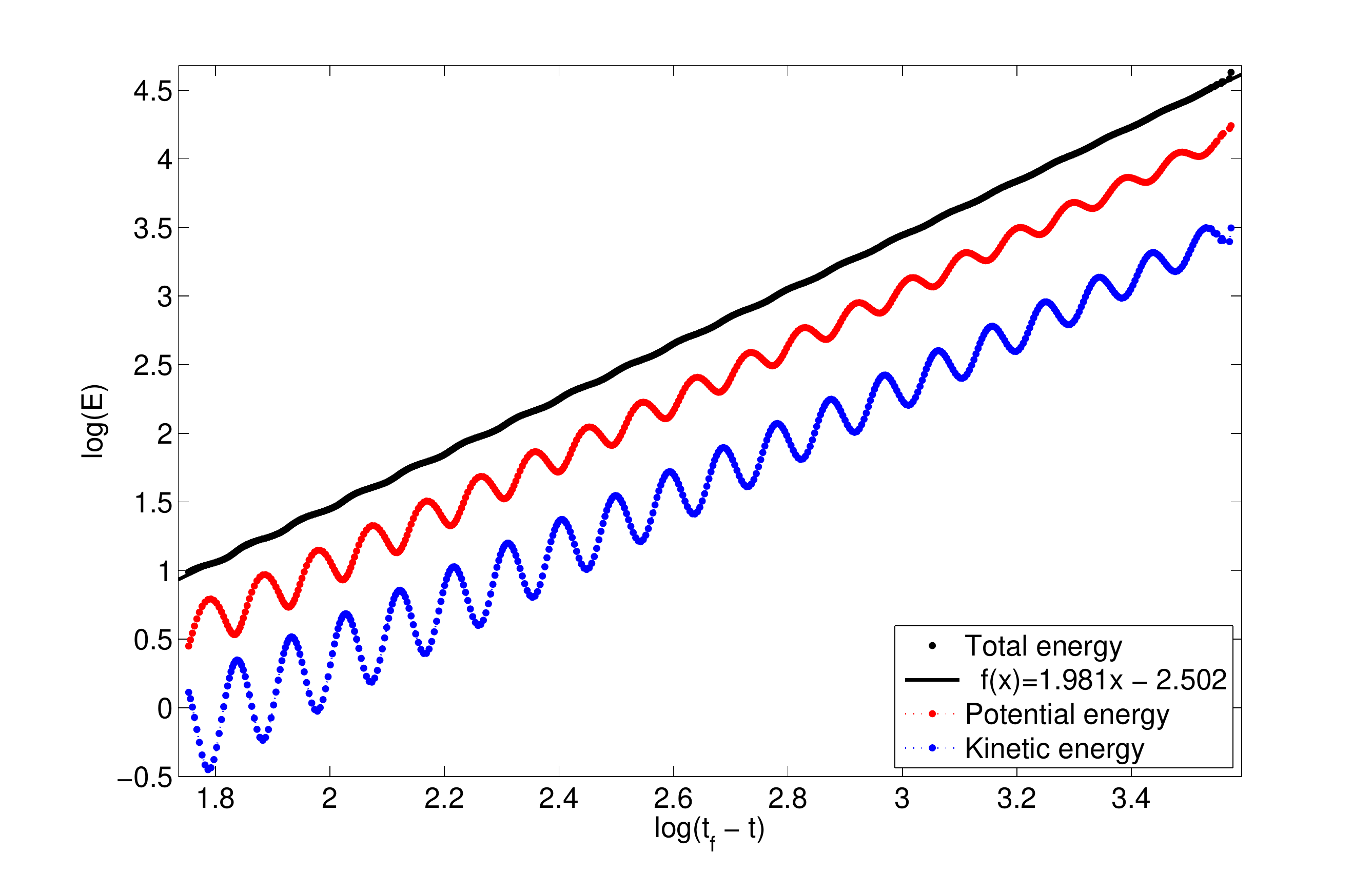}
        \caption{$n=20$ $\gamma=0.7$}
    \end{subfigure}%
    \\
        \begin{subfigure}[t]{0.5\textwidth}
        \centering
        \includegraphics[width=65mm]{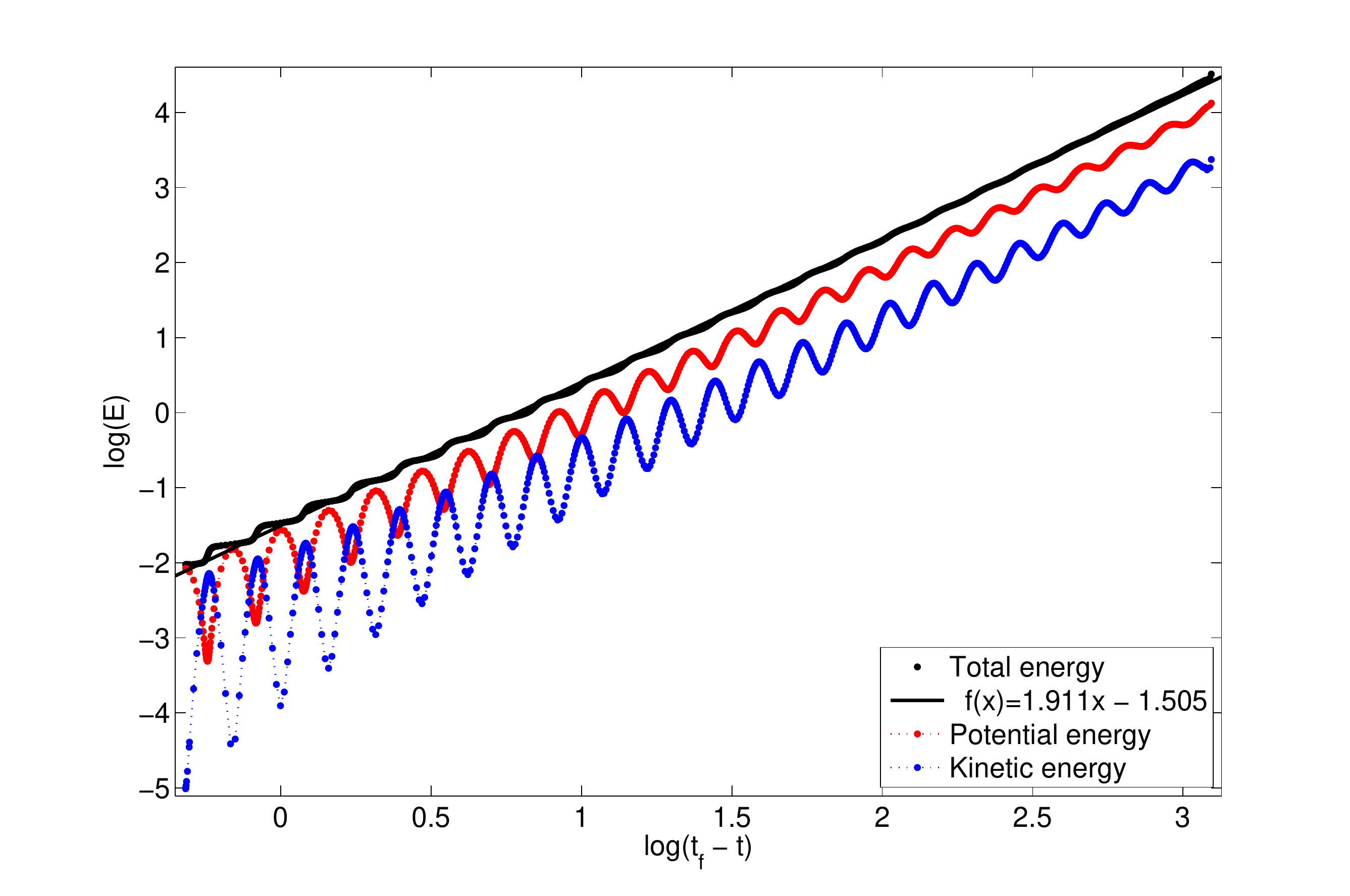}
        \caption{$n=25$ $\gamma=0.5$}
    \end{subfigure}%
    ~ 
    \begin{subfigure}[t]{0.5\textwidth}
        \centering
        \includegraphics[width=65mm]{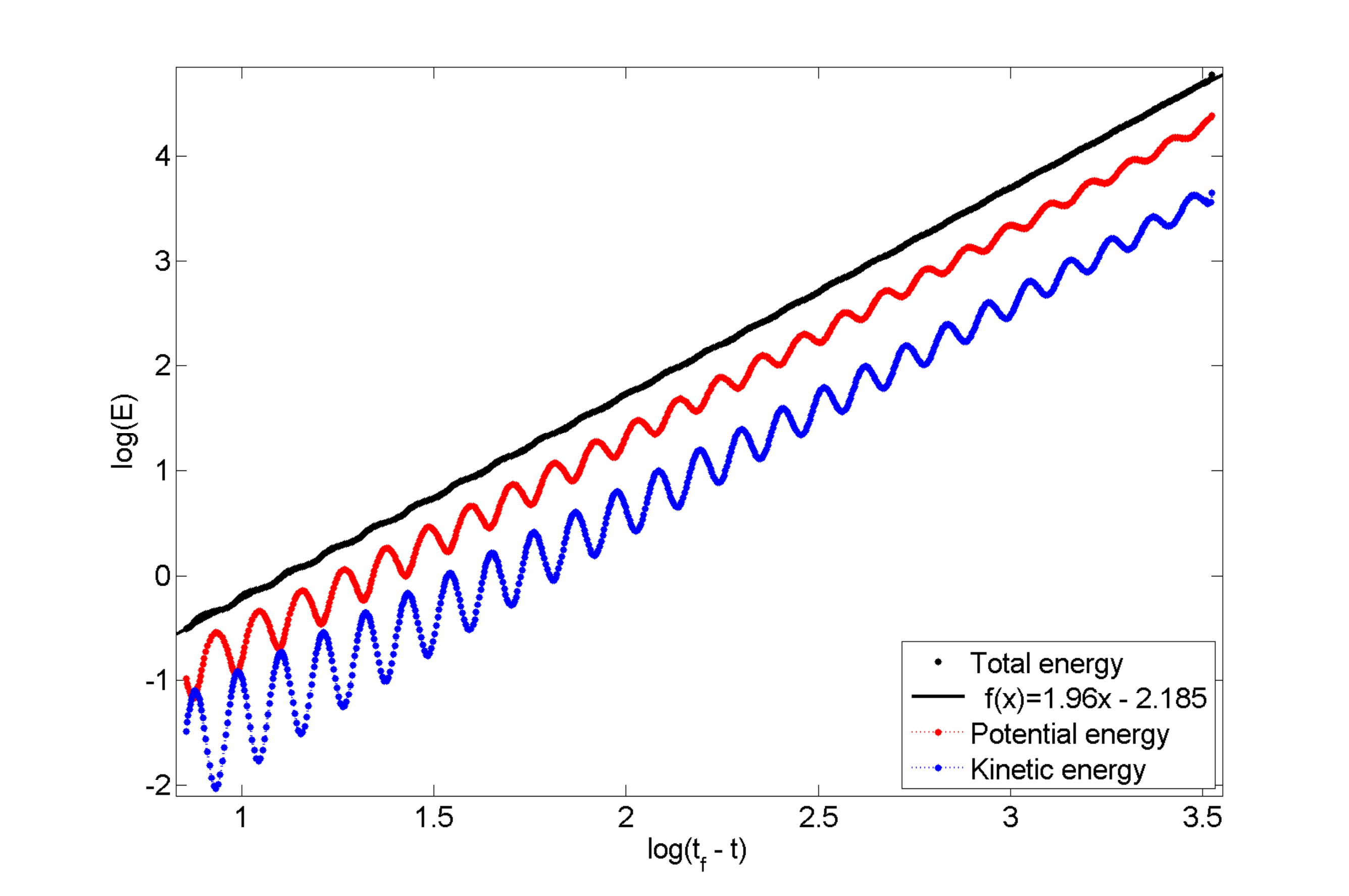}
        \caption{$n=25$ $\gamma=0.6$}
    \end{subfigure}%
    \\
        \begin{subfigure}[t]{0.5\textwidth}
        \centering
        \includegraphics[width=65mm]{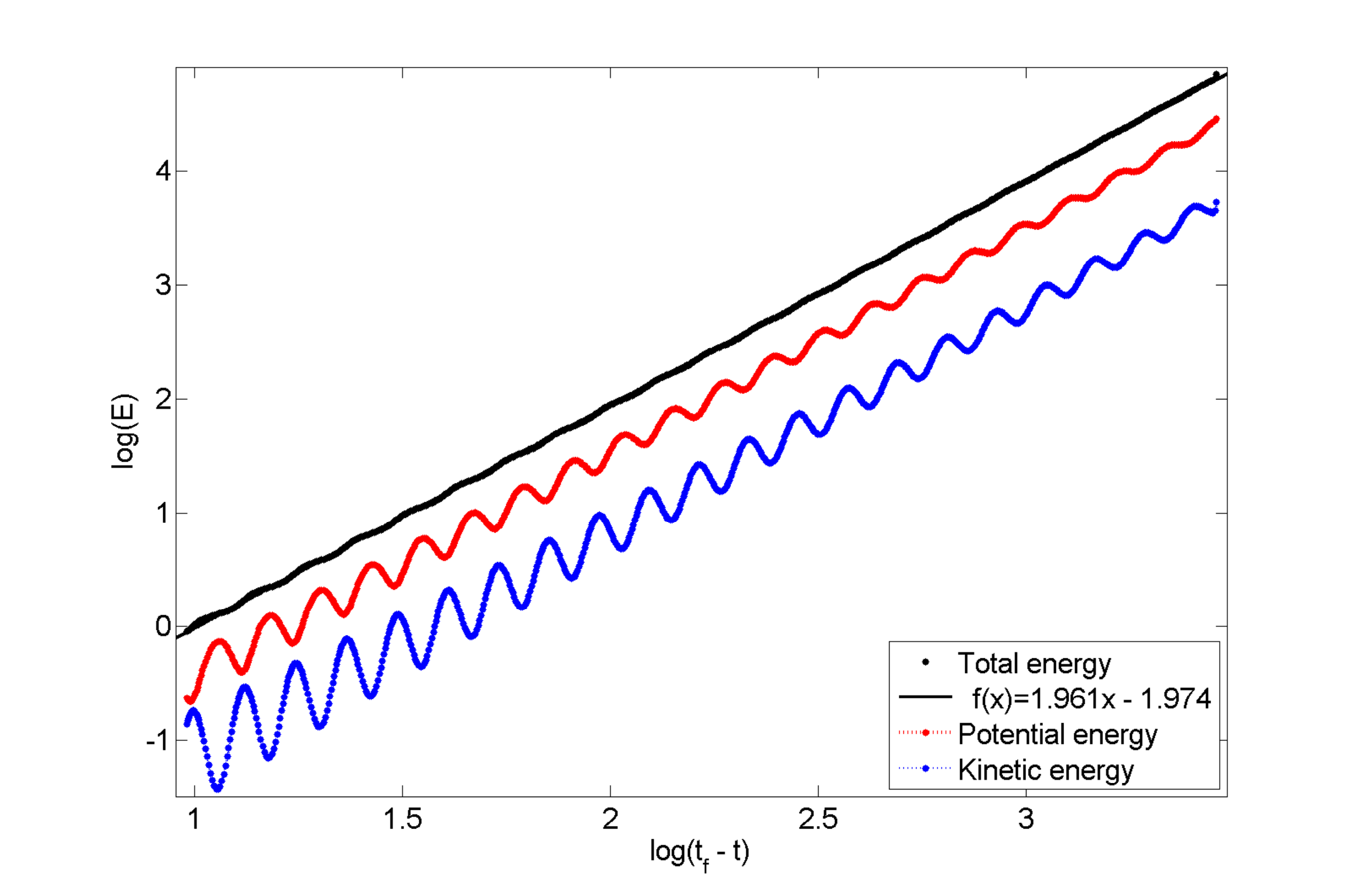}
        \caption{$n=30$ $\gamma=0.5$}
    \end{subfigure}%
    ~ 
    \begin{subfigure}[t]{0.5\textwidth}
        \centering
        \includegraphics[width=65mm]{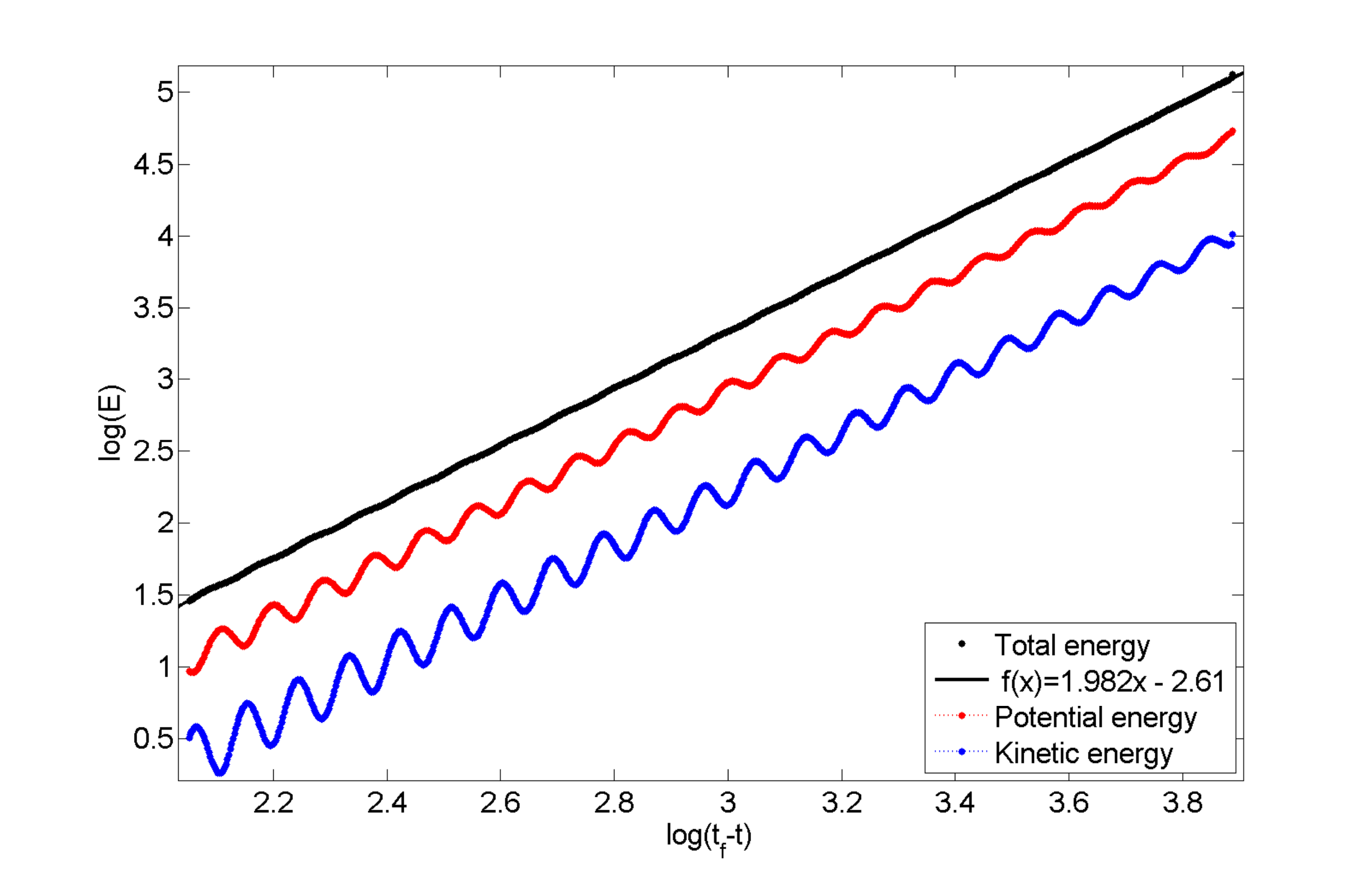}
        \caption{$n=30$ $\gamma=0.6$}
    \end{subfigure}%
    \caption{Log-log plot of energy versus time remaining to singularity obtained by numerical simulation of motion for various values of $n$ and $\gamma$}
    \label{fig:precession}
    \end{figure}

To demonstrate instability, we show results of some direct numerical simulations of the motion in Fig. \ref{fig:precession}. In each case, the initial conditions are slightly perturbed versions of self-similar motion. Precession is reflected by the gradually growing oscillatory components of the kinetic and the potential energy. Hence precession will become more and more dominant as the singularity is approached. The divergence from self-similar motion described above eventually changes the regular order ($0,1,2,...,n-1,0,1,...$) of impact locations and gives rise to more irregular patterns of motion.  Our numerical results are in agreement with the experimental observations of a small wobbling (or precessional) component. Our results question the validity of analysis based on the assumption of precession-free motion. Nevertheless, in the forthcoming section, we argue that the irregularity of the motion does not affect the exponent of energy decay. \\

\subsection{Energy exponent of irregular motion}
We have seen in Sec. \ref{sec:energydissipation} that self-similar motion in the case of inelastic impacts corresponds to energy dissipation exponent $c=2$ in the sense of \eqref{eq:scalinglaw2}. It is easy to show that self-similar motion for partially elastic impacts would lead to the same exponent. Nevertheless, the instability of self-similar motion for $\gamma>0$ means that more irregular motion may occur instead. 
Nevertheless, under the assumtption that 
\eqref{eq:scalinglaw2} is satisfied with some values of $a_1$, $a_2$, and $c$, we are able to prove that the only possible value of $c$ is 2. The proof of this statement is sketched in Appendix \ref{sec:exponent}. The simulation results in Fig. \ref{fig:precession} also indicate
$c\approx 2$. These observations bring us again to the final conclusion that impacts are probably not the dominant mechanism of energy dissipation of Euler's disk.

\section{Modelling irregular support surfaces}

So far we have been investigating the effect of the imperfect shape of the disk itself. Another approach to the question of irregularity is to suppose the supporting surface is not perfectly planar. If a round disk spins on slightly curved terrain, the speedup of spinning will cause liftoff as the disk passes the "bumps" of the underlying surface. This leads to rapid sequences of collisions on the bumps of the surface and energy dissipation. 

\begin{figure}[h]
\begin{center}
\includegraphics[height=4cm]{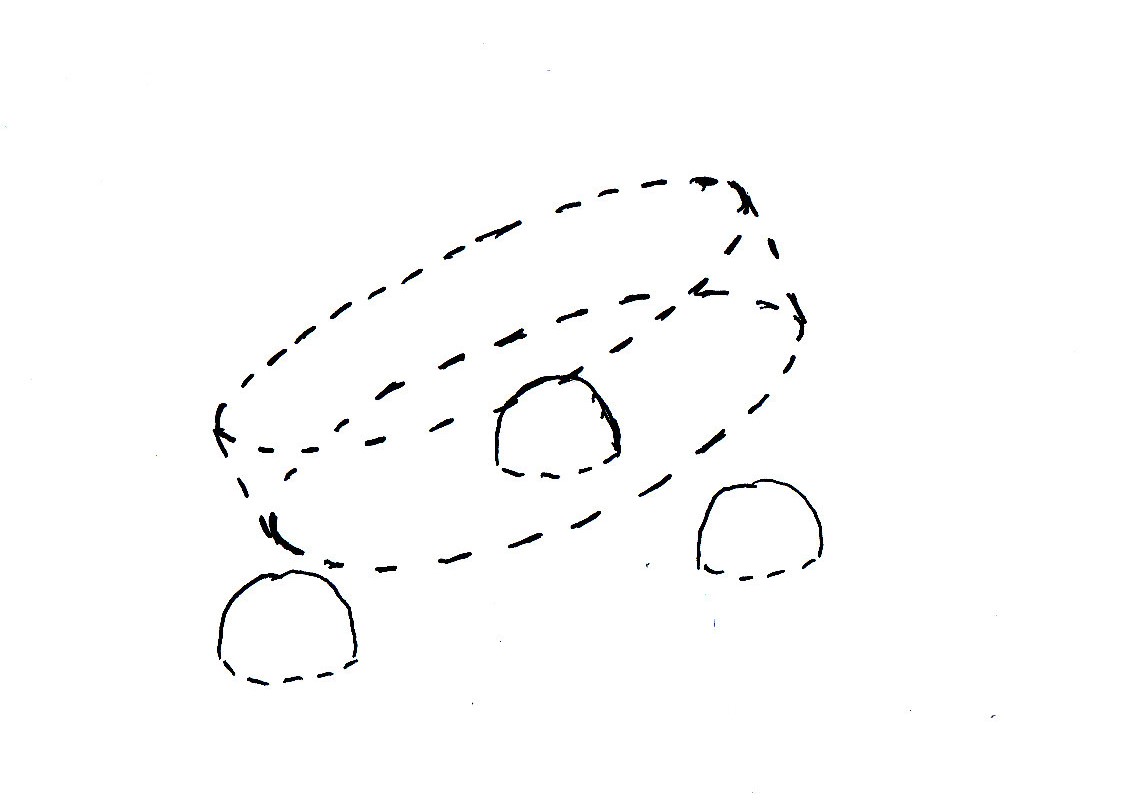}
\caption{Sketch of the theoretical model accounting for irregularity of the support surface}
\label{fig:3bump}
\end{center}
\end{figure}

If two nearly flat (but slightly irregular) rigid surfaces are pushed against each other then they will typically have 3 ponts of contact. Motivated by this simple geometric observation, our conceptual model to study the dynamics of the disk over a slightly irregular terrain will be a rigid disk bouncing on 3 point contacts (Fig. \ref{fig:3bump}). For simplicity, we will assume that these points do not vary over time, and also that they form an equilateral triangle. 
These assumptions allow us to rely on the results of the previous sections (but now we consider the case $n=3$).

Throughout the previous two sections, we focused on self-similar motion, which was analogous to the precession-free motion of round disks. Nevertheless, now we have no reason to restrict our attention to self-similar motion, which makes the analysis more complicated. As a first step, we review results derived by other researchers with respect to the planar motion of a falling rod (which correwponds to $n=2$ and $\omega_x=\omega_z=0$ in our notation). Then we use numerical techniques to explore spatial motion in the $n=3$ case. We will focus on the case of partially elastic impacts $\gamma>0$, since $\gamma=0$ often leads to simulateneous impacts, for which we lack good impact models.

\subsection{Literature review of the falling rod problem}
Consider a rod of length 2 with symmetrical mass-distribution hitting a flat surface. This system correspons to the $n=2$ case of the model of the present paper. Let the mass-distribution parameter $\psi$ be defined as 
$$
\psi=\frac{1-\rho^2}{1+\rho^2}
$$
where $\rho$ is the radius of gyration. 
($\psi=0$ corresponds to mass concentrated at the endpoints, $\psi=1$ to mass concentrated to the center and $\psi=0.5$ to homogenous mass distribution). The planar motion ($\omega_x=\omega_z=0$ throughout the motion) of such a rod  under patrially elastic contacts ($\gamma>0$), \textit{in the absence of gravity} ($g=0$) has been investigated by \cite{goyal1,goyal2}. They found that if
\begin{align}
\psi >\frac{2\gamma^{1/2}}{1+\gamma}
\label{eq:CCcond}
\end{align}
then the motion follows one of the following two patterns depending on the choice of initial conditions: 
\begin{itemize}
\item (complete chatter) - motion converges asymptotically to a regular pattern of motion: the two endpoints of the rod hit the ground in alternating order. Due to the absence of gravity, the generalized velocities remain constant during episodes of free-flight 
After the $j^{th}$ impact, the generalized velocity is $p_j=[\omega_{x,j} \omega_{y,j} v_j]^T$ with
\begin{align}
v_j &=-|v_0|\beta_p^j\\ 
\label{eq:rodendpoint velocities1}\\
\omega_{x,j}&=0\\
|\omega_{y,j}| &= |v_j|\frac{1+\beta_q}{1-\beta_q} 
\label{eq:rodendpoint velocities2}
\end{align}
where
\begin{align}
\beta_p&=\frac{\psi(1+\gamma)+\sqrt{\psi^2(1+\gamma)^2-4\gamma}}{2}
\\
\beta_q&=\frac{2\gamma}{\psi(1+\gamma)+\sqrt{\psi^2(1+\gamma)^2-4\gamma}}
\end{align}

Complete chatter occurs among others if the initial velocity is rotation-free. The exact range of initial conditions corresponding to this scenario are determined in a previous paper by us (Sec. 3 of \cite{baranyai2017zeno}). 
\item (incomplete chatter)
- the rod separates from the ground after a finite number of impacts. This scenario is realized among others for all initial conditions such that the initial velocity of the midpoint of the rod points away from the support surface.
\end{itemize}
If the condition \eqref{eq:CCcond} is not satisfied, then complete chatter becomes impossible and any initial condition leads to incomplete chatter.

If the same rod moves under the effect of gravity (and the gravity points towards the support surface) then its behaviour changes significantly. Incomplete chatter becomes impossible due to energy bounds of the system. The behaviour of the system has been investigated in \cite{or2008hybrid} and \cite{orthesis}, where they find a condition slightly stricter than \eqref{eq:CCcond}: 
\begin{align}
\psi > \frac{\gamma^{1/3}+\gamma^{2/3}}{1+\gamma}
\label{eq:CCcondgravity}
\end{align}
If \eqref{eq:CCcondgravity} is satisfied, then the motion of the rod follows one of the following scenarios:
\begin{itemize}
\item an infinite sequence of impacts occurs at one endpoint (while the other endpoint remains separated). This sequence terminates in finite time, giving rise to sustained contact at one endpoint. Some time later, the other endpoint hits the ground giving rise to a simultaneous impact at the two endpoints. Our impact model is not applicable to this scenario. Those authors make some assumptions about simultaneous impacts, to infer that the object stops immediately after the simultaneous impact.
\item after some initial transient, the two endpoints hit the ground in alternating order. The flight times shrink with a fast decay rate such that gravity has less and less influence on the motion. In addition, motion converges to the gravity-free, complete chattering sequence described above.
\end{itemize}
If we have gravity but \eqref{eq:CCcondgravity} is not satisfied, then it is easy to see that the rod may not separate from the ground (due to bounded potential energy) and the influence of gravity may not become negligible (since this would imply separation according to \cite{goyal1},\cite{goyal2}). These two observations imply that the rod undergoes some kind of motion, in which the effect of gravity remains significant. This form of motion has not been investigated in detail.

\subsection{Energy exponent of the falling rod}
\label{sec:rodexponent}
From the point of view of our analysis the most interesting task is showing that the power-law relationship \eqref{eq:scalinglaw2} applies for the rod and finding the exponent. We have argued previously that for any form of motion in which gravity plays significant role, the only possible value of the exponent is $c=2$. This result applies to the rod as well. Nevertheless it is not applicable to any of the two possible scenarios if \eqref{eq:CCcondgravity} is true.

The first scenario leads to simultaneous impact, where the lack of a good model prevents us from detailed analysis. Nevertheless, if we adopt the hypothesis of other researchers that the rod stops immediately, then this scenario loosely corresponds to $c=0$. 

The second scenario is convergence to gravity-free  complete chattering motion. Interestingly, \eqref{eq:rodendpoint velocities1}-\eqref{eq:rodendpoint velocities2} show that this motion obeys the condition \eqref{eq:pscaling} of self-similarity with $\beta=\beta_p$, i.e. we have
\begin{align}
p_i&=\beta_p^iR^ip_0
\label{eq:pscaling-rod}
\end{align}
Nevertheless \eqref{eq:qscaling} is not satisfied. Instead of that, one can easily show that it satisfies another condition 
\begin{align}
q_i&=\beta_q^{i}R^iq_0
\label{eq:qscaling-rod}
\end{align}
We also  note that $\beta_p$ and $\beta_q$ are two positive eigenvalues of the matrix $R^{-1}U_0$ with $n=2$ (see Appendix \ref{sec:trinog}). 

The scaling properties
 imply that 
the durations of the episodes of free flight decrease exponentially as
\begin{align}
\tau_j=\tau_0\left[\beta_q/\beta_p
\right]^{j}
\label{eq:rodtimes}
\end{align}
and thus impacts accumulate at a finite-time singularity after which both endpoints are in sustained contact with the support surface.  
  
Eq. \eqref{eq:pscaling-rod} also implies that the kinetic energy of the rod decreases exponentially as 
\begin{align}
E_j=E_0\beta_p^{2j}
\label{eq:rod-energylaw}
\end{align}
whereas the system has no potential energy du to the absence of gravity. We can now repeat the calculation presented in Sec. \ref{sec:energydissipation} but with \eqref{eq:rodtimes} instead of \eqref{eq:tauscaling} and with \eqref{eq:rod-energylaw} instead of \eqref{eq:Ei}. This way, we find that \eqref{eq:scalinglaw2} is satisfied with exponent
\begin{align}
c_{lin}=
\frac{2\log\beta_p}{\log(\beta_q/\beta_p)}\label{eq:clin}
\end{align}
\begin{figure}
\begin{center}
\includegraphics[width=10cm]{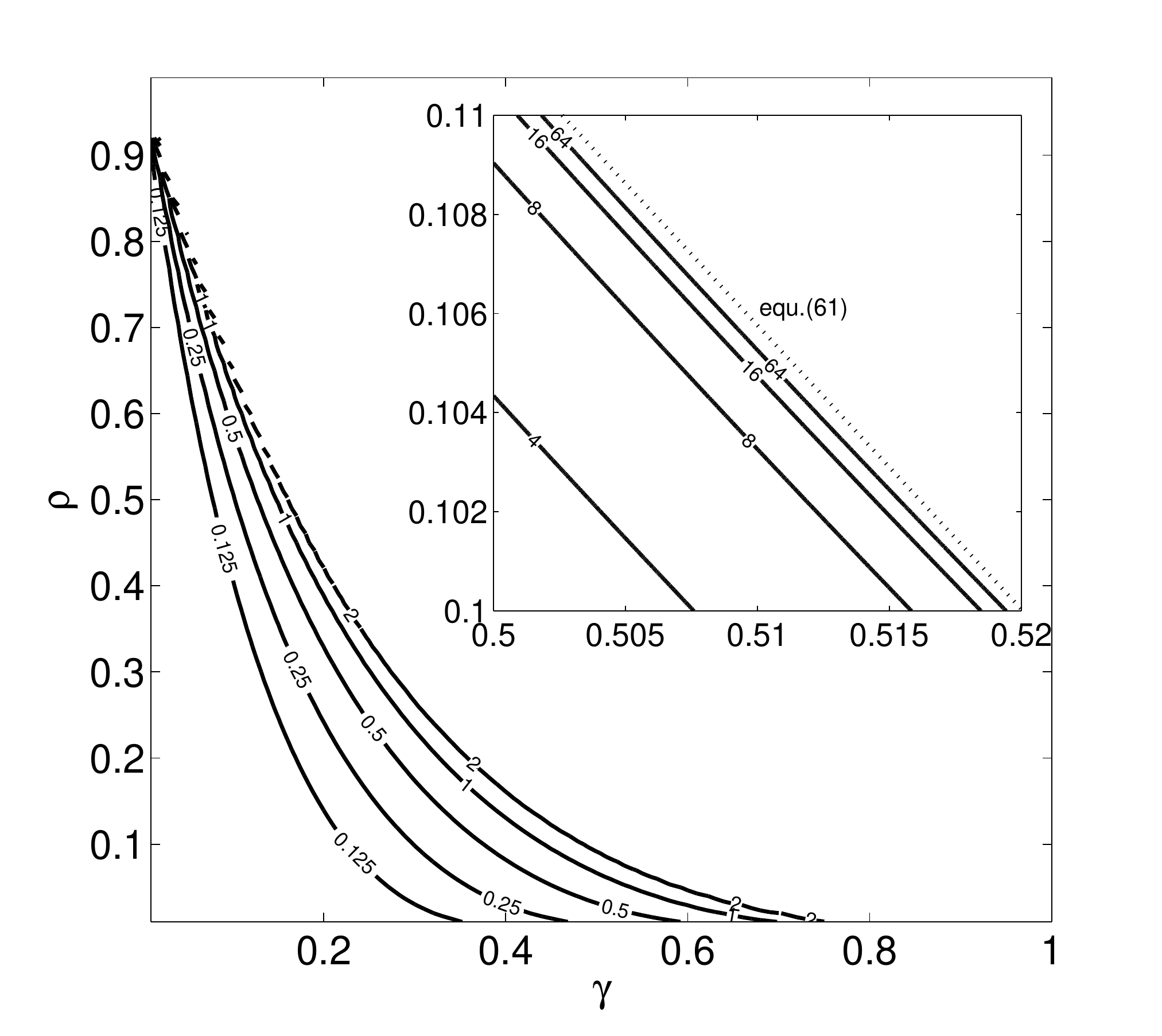}
\caption{Exponents of the falling rod when gravity is negligible.}\label{fig:rodcoeff}
\end{center}
\end{figure}

The actual values of $c_{lin}$ in terms of $\psi$ and $\gamma$ are shown in Fig \ref{fig:rodcoeff}. We notice without detailed proof that $c_{lin}$ goes to infinity if we approach boundaries of the region determined by \eqref{eq:CCcond}; it converges to 0 if either $\gamma\rightarrow 0$ or $\rho\rightarrow 0$; furthermore  it is equal to 2 whenever there is equality in \eqref{eq:CCcondgravity}.

Hence, we conclude that the energy exponent of the rod can be arbitrarily close to 0. Next we will examine the motion of a bouncing triangle. An analogous result in that case would mean that energy absorption due to impacts may become dominant over all previously investigated dissipation mechanisms.

\subsection{Energy exponent of a bouncing disk with 3 contact points}

We have seen that for low $\gamma$ and $\rho$ values, a rod either undergoes motion cumulating in a simultaneous impact, or complete chatter if gravity is absent. In a previous paper \cite{baranyai2017zeno}, we have demonstrated by using analytical and numerical tools that a flat polygonal disk with multiple contact points behaves in the same way in the absence of gravity. In that paper, we have found that 
the motion always leads to an infinite sequence of impacts at two vertices (terminating in finite time), followed by a simultaneous impact if it satisfies the following criterion:
\begin{definition}
A system possesses the partial complete chattering (PCC) property, if the matrix $U_0RU_0R^{-1}$ has no complex eigenvalues, that is:
\begin{align}
URUR^{-1}p=\nu p \implies \nu \in \mathbb{R} 
\label{eq:PCCtriangle}
\end{align}
\end{definition}

Furthermore it is also shown there that a polygonal body in a gravitation-free environment undergoes complete chatter resulting in simultaneous contact at all vertices if it satisfies the following criterion:
\begin{definition}A system possesses the complete chattering (CC) property, if the dominant eigenvalue of the matrix $U_0R^{-1}$ is real and positive, i.e. if
\begin{align}
\sup(\{ \lambda_i\})=sup(\{\abs{\lambda_i}\}) \text{ and } sup(\{ 
\lambda_i\})\in \mathbb{R},
\label{eq:CCtriangle}
\end{align} 
where $\lambda_i$ ($i=1,2,3$) are the eigenvalues of matrix $R^{-1}U_0$.
\end{definition}

Due to the difficulties of modelling simultaneous impacts if both \ref{eq:PCCtriangle} and \ref{eq:CCtriangle} are satisfied, we will focus on  systems with the CC but without the PCC property. We show in Appendix \ref{sec:trinog} that a triangle also has a self-similar motion  in the sense of \eqref{eq:pscaling-rod} and \eqref{eq:qscaling-rod}. The corresponding energy exponent is again of the form \eqref{eq:clin} where now $\beta_p,\beta_q$ denote two positive eigenvalues of $R^{-1}U_0$ with $n=3$.
Nevertheless we also find by numerical analysis of a Jacobian that this motion is unstable unlike in the case of the rod (details omitted).\\

\begin{figure}[h]
\begin{center}
\includegraphics[width=15cm]{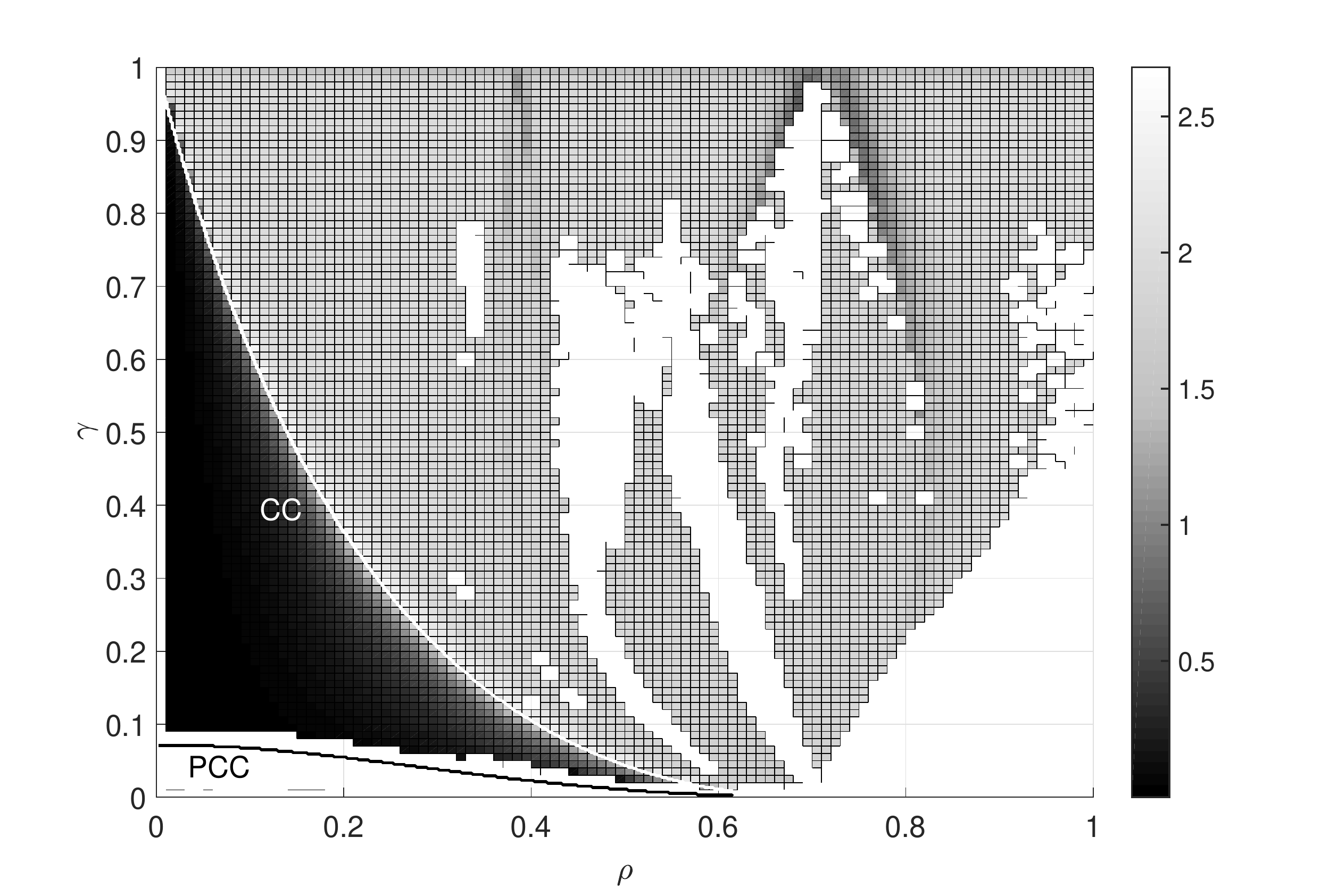}
\caption{Numerically obtained exponents in the case of $n=3$. Lack of values means simultaneous impacts. The curves labelled by $CC$ and $PCC$ are the boundaries of parameter regimes with the CC and PCC properties.}
\label{fig:numsim}
\end{center}
\end{figure}

\begin{figure}[h]
\begin{center}
\includegraphics[width=15cm]{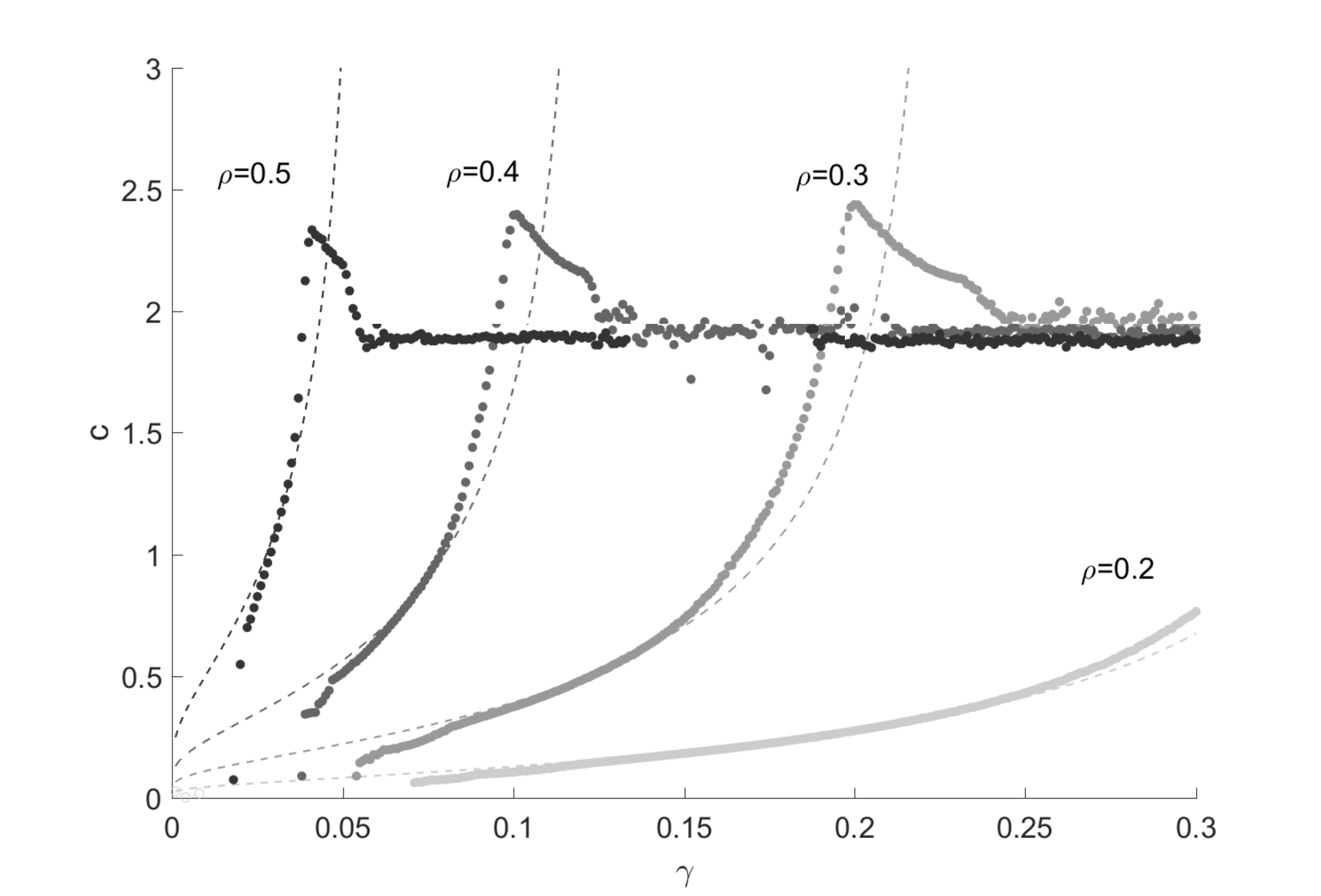}
\caption{Sections of Fig. \ref{fig:numsim} (dots) with predictions of \ref{eq:clin} (dashed curves).}
\label{fig:numsimsec}
\end{center}
\end{figure}

To account for irregular motion and to see if the role of gravity becomes negligible during the final stage of the motion,
we conducted a series of numerical simulations with $g=1$, in which $\gamma\in[0 \ 1]$ and $\rho\in[0 \ 1]$ were varied systematically over a rectangular grid. At each point of the grid 100 simulations were conducted. In each simulation, the value of $c$ was estimated by linear regression over the logarithm of energy vs. time. The simulation often terminated due to simultaneous impacts for which we lack a good model. In these cases no value has been recorded for $c$. For each simulation, the initial values of $p$ and $q$  were given as $p=[0.1 \chi_1; 0.1 \chi_2; -1]^T$ and $q=[0.001 \chi_3; 0.001 \chi_4; 1]^T$ where the variables $\chi_1 ... \chi_4$ were drawn randomly from a uniform distribution over the interval $[-0.5 \ 0.5]$.
 The result of this process is summarized in Fig. \ref{fig:numsim}, where the average values of the 100 simulations are shown. Where no value is provided, all 100 simulations ended in simultaneous collisions. The theoretical boundaries of \ref{eq:PCCtriangle} and \ref{eq:CCtriangle} are also provided. In Fig. \ref{fig:numsimsec} we also show the estimated exponents as a function of $\gamma$ for four values of $\rho$ with results of the simulations as a comparison.

Our results suggest the following:
\begin{itemize}
\item if \eqref{eq:PCCtriangle} (and thus \eqref{eq:CCtriangle}) is satisfied, then the motion includes a simultaneous impact and our model is not capable of predicting the final outcome of the motion.
\item if \eqref{eq:CCtriangle} is satisfied while 
\eqref{eq:PCCtriangle} is not; then the object undergoes irregular motion, in which the effect of gravity becomes negligible. The values of $c$ estimated from simulations are close to the values $c_{lin}$ predicted by \ref{eq:clin} and are mostly below 2 (Fig. \ref{fig:numsimsec}).
\item if \eqref{eq:CCtriangle} is not satisfied, then the effect of gravity remains significant throughout the motion, and we have $c\approx 2$. The simulation often but not always terminates due to simultaneous impacts. 
\end{itemize}
The estimated exponents are significantly above 2 near the boundary of the region \eqref{eq:CCtriangle}. We believe that this is a numerical artefact, which was caused by the finite lengths of simulations.

Our results show that the exponent may be very low. Unfortunately, we were not able to find low values of $c$ in the case of $\rho=0.5$, which corresponds to Euler's disk with homogeneous mass distribution. This shortcoming of our results is due to the emergence of simultaneous impacts. Nevertheless, lower values of $\rho$ often lead to low values of $c$, in which case energy dissipation due to impacts becomes the dominent dissipation mechanism.

\section{Conclusions}

We have examined in this paper how impacts initiated by the geometric imperefection of a spinning disk affect the singularity at the end of its motion. Two types of imperfection have been analyzed: the imperfect rotational symmetry of the disk was modelled by a polygonal profile, and the imperfections of the flat underlying surface were modelled by considering a rigid object with 3 point contacts to the ground.

In order to simplify the analysis, we  used impacts with constant coefficient of restitution and frictionless contact interactions (during continuous motion and impacts). In addition, a discrete rotational symmetry has been assumed in both models. 

In accordance with the majority of the previous works we focused on the rate of energy dissipation shortly before the singularity. This question is usually investigated under the assumption that the disk undergoes precession-free motion, i.e. that the slope of the spinning disk decreases monotonically. 
In the framework of our first model, we defined the concept of self-similar motion and demonstrated that it is analogous to the precession-free spinning of a perfectly round, dissipation-free disk. Nevertheless self-similar motion naturally  incorporates energy dissipation via impacts.  We demonstrated that the energy profiles generated by impacts have an exponent $c=2$, which is higher than the exponent of many other energy dissipation mechanisms. This means that impacts are not the dominant dissipation mechanism during the last phase of motion. Moreover, we also showed that self-similar motion may be destabilized by impacts, which provides a new explanation of the significant precessional component of motion often found  in physical experiments. 

Our second model revealed that the lack of perfect flatness of the underlying ground has a much more subtle effect. In many cases, the motion of our conceptual model includes simultaneous impacts at multiple points, for which the lack of an impact model prevented us from detailed analysis. Nevertheless, we also found a range of model parameters and intial conditions, for which simultaneous impacts do not occur, moreover impacts absorb the energy of the disk at an increasing rate and the exponent of the corresponding  energy profile may take any value in the range $c\in [0 \ 2]$. This finding is surprising since any other previously examined dissipation mechanism revealed exponents $c=4/9$ or higher. Hence our results suggest that energy absorption during the last phase of motion of a spinning disk may be dominated by impacts  in the case of a relatively low coefficient of restitution and radius of gyration. This result remains partial since we do not examine the transition from disk-like behaviour to bouncing motion on 3 point contacts and thus we do not demonstrate that the second type of motion is initiated with appropriate initial conditions.

As \cite{leine2009experimental} pointed out, we may find different dominant energy dissipation mechanisms depending on the time-scale chosen. The main conclusion of that paper is that rolling friction (with dissipation exponent 0.5) is dominant over the time scale of seconds but other effects (with lower exponents) may become dominant during the last few milli-seconds. We now provide a rough estimation of the time scale over which impacts may become dominant.  Assume that the actual shape of the underlying surface deviates from its ideal, planned shape by $\Delta\approx 10^{-4...-5}$ m. and that the diameter of the disk is $d=10^{-1}$ m. Then, the triangle model yields a reasonable description of the motion, when the inclination angle drops significantly below $\alpha\approx\Delta/d= 10^{-3...-4}$ rad. According to the experimental results of Leine (Fig. 6 of \cite{leine2009experimental}), the inclination angle drops below this critical value during the last $0.1...1$ second of the motion. Hence, we conclude that imperfections may affect energy dissipation during the last few tenths of second. This estimation confirms with the observation that the last second is accompanied by a strong rattling noise that has been attributed to impacts by some authors. We also believe that if the system has the CC property, then the effect of impacts quickly becomes dominant, once the first impacts occur. This prediction is based on the observation that our triangle model undergoes intensive energy dissipation with the inclination angle dropping from $10^{-4}$ rad to zero within only $\approx 10^{-5}$ sec.

The most important restriction of our analysis is the assumption of ideal rigidity. Clearly, when a thin disk undergoes rapid sequences of impacts, then its motion may be accompanied by significant elastic vibrations. When the inclination angle of the disk is small, then even small-amplitude vibrations have a strong effect on the locations and intensities of impacts between the object and the underlying ground. The analysis of energy dissipation via impacts in a model including elastic deformations is one of the possible directions of further research.

\appendix

\section{Energy dissipation exponent of an $n$-gon}\label{sec:exponent}
Here we demonstrate that the only possible value of the exponent $c$ is 2.

The proof will use the following simple property of impacts with a fixed value $0<\gamma<1$ of the restitution coeffiecient. Let $\Delta E<0$ denote the energy dissipation during an impact and let $\Delta p$ denote the velocity jump vector during the same impact. Then for any given finite-sized system with non-singular mass matrix, there are constants $\kappa_1,\kappa_2>0$ such that every impact satisfies 

\begin{align}
\kappa_1|\Delta p|<\sqrt{-\Delta E}<\kappa_2|\Delta p|
\label{eq:impactdissipation}
\end{align}

Now, let us assume first that \eqref{eq:scalinglaw2} is satisfied with $c<2$. Then for $t_f-t<<1$, the bounds \eqref{eq:scalinglaw2} imply that 
$$
\sqrt{E(t)}<<t_f-t
$$
Together with \eqref{eq:impactdissipation}, this result means that the total variation of $p$ due to impacts during the time interval $(t,t_f)$ is much smaller than $t_f-t$. In other words the effect of impacts on $p$ is negligibly small in comparison with the effect of gravity and contact forces during continuous motion. We must have $p(t_f)=q(t_f)=0$ at the time of the singularity, which means that the center of mass is on the ground and it has zero velocity. Nevertheless during continuous motion, the vertical acceleration of the center of mass is always negative by \eqref{eq:acc2contact},\eqref{eq:acc2free}. Hence the center of mass must penetrate into the ground shortly before the singularity, which is contradiction.

Second, let us examine the possibility of $c>2$ in \eqref{eq:scalinglaw2}. Similarly to the case of $c<2$ we can now conclude that if $t$ is very close to $t_f$, then the effect of continuous motion on $p$ during the time interval $(t,t_f)$ is negligible in comparison the effect of impacts. The bouncing motion of polygonal objects in the absence of gravity has been investigated by the paper \cite{baranyai2017zeno}. Our results show that without the effect of gravity, a regular $n$-gon shaped object will always separate from the ground after a finite number of impacts provided that the radius of gyration is above a certain threshold $\rho_0(n,\gamma)$, which depends on $n$ and $\gamma$. For high values of $n$, this threshold becomes very low, and thus a homogenous disk may not undergo finite-time singularity with infinitely many impacts. This completes our proof that the only possible value of the energy dissipation exponent is $c=2$.

\section{Explicit expressions for $\bar{C}$\label{sec.barC} }
The map $\bar{C}$ is composed of three maps. The first one is a linear map corresponding to impact at vertex $0$:
$$
\begin{bmatrix}
\bar{p}_i\\\bar{q}_i
\end{bmatrix}
\rightarrow
\begin{bmatrix}
p_{ia}\\q_{ia}
\end{bmatrix} :=
\begin{bmatrix}
U_0\bar{p}_i\\\bar{q}_i
\end{bmatrix}
$$
where $U_0$ is given by \eqref{eq:impactequ}. The second map corresponds to continuous motion starting at this impact and ending when vertex $1$ hits the ground:
$$
\begin{bmatrix}
p_{ia}\\q_{ia}
\end{bmatrix}
\rightarrow
\begin{bmatrix}
p_{ib}\\q_{ib}
\end{bmatrix} :=
\begin{bmatrix}
p_{ia}
+a_{contact}\tau_i
\\
q_{ia}
+p_{ia}\tau
+a_{contact}\tau_i^2/2
\end{bmatrix}
$$
where $a_{contact}$ is given by \eqref{eq:pacccontact} and $\tau_i\geq0$ is the duration of motion, which can be determined from the requirement that vertex 1 hits the ground at the end of the motion, i.e.
$$
f_1^T(q_{ia}
+p_{ia}\tau_i
+a_{contact}\tau_i^2/2)=0
$$
This is a second-order equation for $\tau_i$ with exactly one non-negative root.
Finally, the third component of $\bar{C}$ is the transformation $T_1$:
$$
\begin{bmatrix}
p_{ib}\\q_{ib}
\end{bmatrix} 
\rightarrow
\begin{bmatrix}
\bar{p}_{i+1}\\\bar{q}_{i+1}
\end{bmatrix} 
:=T_1
\left(
\begin{bmatrix}
p_{ib}\\q_{ib}
\end{bmatrix}
\right)
$$
These components together yield the expressions \eqref{eq:barCexpression01}, \eqref{eq:barCexpression02}.

\section{Self-similar motion in the absence of gravity}\label{sec:trinog}
Here it will be shown for arbitrary value of $n$, that the conditions \eqref{eq:pscaling-rod} and  \eqref{eq:qscaling-rod} are together equivalent of requiring that $\beta_p$ and $\beta_q$ are both positive, real eigenvalues of $R^{-1}U_0$. The equivalence is demonstrated in to steps
 
\emph{Step 1: Self-similarity implies that $\beta_p$, $\beta_q$ are eigenvalues:}\\
The self-similarity conditions \eqref{eq:pscaling-rod} - \eqref{eq:qscaling-rod} can be expanded as
\begin{align}
U_0 p_0&=R\beta_p p_0
\label{eq:selfsim-p-nog}\\
q_0+U_0 \tau_0 p_0&=R\beta_q q_0
\label{eq:selfsim-q-nog}
\end{align}
The first condition can be rearranged as
\begin{align}
R^{-1}U_0 p_0=\beta_p p_0, \label{eq:segedapp2}
\end{align}
i.e, $\beta_p$ is an eigenvalue and $p_0$ is the corresponding eigenvector of $R^{-1}U_0$. The second condition can be rearranged to obtain either one of the following two formulae:
\begin{align}
R^{-1} q_0+R^{-1}U_0p_0\tau_0=\beta_q q_0\label{eq:segedapp3}\\
R^{-1} q_0=\beta_q q_0 -R^{-1}U_0p_0\tau_0, \label{eq:segedapp}
\end{align}
which will be used below. Also, since $U_0$ represents an impact at vertex 0, we have $f_0^Tx=0 \implies U_0x=x$ for any vector $x$. We can apply this identity to $q_0$, since it has been defined as an impacting configuration, i.e. $f_0^Tq_0=0$. Thus
\begin{align}
U_0 q_0=q_0.\label{eq:u0q0}
\end{align}
Now we write a simple identity and rearrange it in several steps in which the equations \eqref{eq:segedapp3},\eqref{eq:u0q0}, \eqref{eq:segedapp}, and \eqref{eq:segedapp2} are used respectively.
\begin{align}
(R^{-1}U_0)((\beta_p-\beta_q)q_0+\beta_p \tau_0 p_0)&=\beta_p(R^{-1}U_0)(q_0+\tau_0 p_0)-\beta_q(R^{-1}U_0) q_0\\
&=\beta_p\beta_q q_0-\beta_q (R^{-1}U_0)q_0\\
&=\beta_q(\beta_p q_0-R^{-1} q_0)\\
&=\beta_q(\beta_p q_0+R^{-1}U_0 p_0 \tau_0-\beta_q q_0)\\
&=\beta_q((\beta_p-\beta_q)q_0+\tau_0 \beta_p p_0)
\end{align} 
Thus,
\begin{align}
(R^{-1}U_0)((\beta_p-\beta_q)q_0+\beta_p \tau_0 p_0)=\beta_q((\beta_p-\beta_q)q_0+\beta_p \tau_0 p_0)
\end{align}

which means that $\beta_q$ is also an eigenvalue.
\\
\emph{Step 2: Each pair of distinct eigenvalues corresponds to self-similar motion}\\
Let us consider a pair of eigenvalues $\beta_p$ and $\beta_q$ with the corresponding eigenvectors being $p_0$ and $s$. Then, the self-similarity condition \eqref{eq:pscaling-rod} is trivially satisfied. 

Next, we seek a vector
\begin{align}
q_0=\alpha s+\delta p_0
\label{eq:kombinacio}
\end{align} 
such that  \eqref{eq:ftq0} and \eqref{eq:selfsim-q-nog} hold. To this end we plug \eqref{eq:kombinacio} into \eqref{eq:selfsim-q-nog} and perform some algebraic manipulation:
\begin{align}
\alpha s+\delta p_0+U_0 \tau_0 p_0&=R\beta_q (\alpha s+\delta p_0)\\
R^{-1}(\alpha s+ \delta p_0)+R^{-1}U_0\tau_0 p_0&=\beta_q(\alpha s+ \delta p_0)\\
R^{-1}U_0(\alpha s+ \delta p_0 +\tau_0 p_0)&=\beta_q(\alpha s+ \delta p_0)\\
\alpha \beta_q s+\delta \beta_p p_0 +\tau_0 \beta_p p_0&=\beta_q \alpha s +\beta_q \delta p_0\\
(\delta \beta_p+\tau_0 \beta_p)p_0&=\beta_q \delta p_0\\
\delta \beta_p+\tau_0 \beta_p&=\beta_q \delta\\
\delta&=\beta_p \tau_0 / (\beta_q-\beta_p)
\end{align}
which is a valid solution whenever $\beta_p \neq \beta_q$.

The condition $f^Tq_0$ yields:
\begin{align}
f^T(\alpha s+\delta p_0)=0\\
\alpha=\delta f_0^Tp/f^Ts
\end{align}
This formula is valid if $f^Ts\neq0$, holds, which is provably always true.Hence we have found values of $\delta$, $\alpha$ such that that the resulting value of $q_0$ given by \eqref{eq:kombinacio} satisfies all conditions.




\bibliographystyle{ieeetr}

\bibliography{eulerdisk}

\end{document}